\documentclass[aps,pre,twocolumn,superscriptaddress,longbibliography,floatfix]{revtex4-1}

\usepackage[utf8]{inputenc}
\usepackage[english]{babel}
\usepackage[T1]{fontenc}
\usepackage{lmodern}

\usepackage{amsmath,amsthm,amsfonts,amssymb,bm,amssymb}

\usepackage[colorlinks={true}, citecolor={blue}, filecolor={blue}, linkcolor={blue}, urlcolor={blue}]{hyperref}

\usepackage{graphicx}
\usepackage[caption=false]{subfig}
\graphicspath{{figs/}} 

\usepackage{microtype}
\usepackage{soul}

\newcommand{\Tr}{\operatorname{Tr}}

\newcommand{\kB}{k_\mathrm{B}}

\begin{document}

\title{Topological and finite size effects in a Kitaev Chain Heat Engine}

\author{Elif Yunt}
\email{eyunt@ku.edu.tr}
\affiliation{Department of Physics, Ko\c{c} University, 34450 Sariyer, Istanbul TURKEY}

\author{Mojde Fadaie}
\email{mfadaei@ku.edu.tr}
\affiliation{Department of Physics, Ko\c{c} University, 34450 Sariyer, Istanbul TURKEY}

\author{\"{O}zg\"{u}r E. M\"{u}stecapl{\i}o\u{g}lu}
\email{omustecap@ku.edu.tr}
\affiliation{Department of Physics, Ko\c{c} University, 34450 Sariyer, Istanbul TURKEY}


\begin{abstract}

	We investigate a heat engine with a finite length Kitaev chain in an Otto cycle. Finite size effects are taken into account using method of Hill's nanothermodynamics as well as using the method of temperature dependent energy levels. We distinguish the bulk and boundary contributions to the efficiency and work output of Kitaev chain engine and identify them as non-Otto heat engine and refrigerator cycles, respectively. Possibility of separately running Otto engine cycles associated with the bulk and the whole system, and an Otto refrigerator at the boundary is pointed out. It is found that the critical point of the topological phase coincides with the extremum of the efficiency and the work output of the bulk and the total Otto engine.

\end{abstract}

\maketitle
\section{Introduction}
 
Thermodynamic methods to describe energy processes are strictly applicable only to macroscopically large objects. Progress in engineering ever smaller energy devices makes the question of how to extend the scope of thermodynamics to finite-size structures increasingly significant. For macromolecules, such as polymers, 
a formulation of thermodynamics has been proposed in the early 1960s by T.~L.~Hill~\cite{hillbook,doi:10.1002/ijch.196500008},  and called as nanothermodynamics~\cite{e17010052,hill_perspective_2001,doi:10.1021/nl010027w} whose general principles can be applied to other finite-size systems, too. In particular, phases of topological insulators have been studied using nanothermodynamic approach  ~\cite{quelle,universality} and relation to Uhlmann Phase approaches~\cite{delgado1,delgado2} are pointed out; while nanothermodynamics predict a negligible boundary effect on the anomalous heat capacity in topological Kondo insulators~\cite{broeke_thermodynamic_2018}. Our recent studies suggest that gap-closing topological phase transitions can be probed using work output of quantum heat engines~\cite{ PhysRevE.98.052124}. Here we ask if similar conclusions can be reached in the presence of finite-size effects and examine a nanothermodynamic heat engine of a finite-length Kitaev chain ~\cite{ Kitaev_2001}.

Kitaev chain can exhibit a topological phase transition (TPT) from a trivial to a topological insulator phase in which it hosts topological excitations of zero-energy Majorana modes localized at the edges; whose nonabelian braiding statistics are appealing for topological quantum computation~\cite{KITAEV20032}. Due to the challenges of their direct observation, indirect probing schemes have been discussed ~\cite{Alicea_2012}. An intriguing proposal is to utilize stroboscopic heat current in a time periodic modulated spin chain to detect signatures of Floquet-Majorana modes ~\cite{PhysRevB.96.125144}. Our scheme addresses a static Kitaev chain and makes a rigorous treatment of heat transfer into the system by taking into account finite-length effects using nanothermodynamic methods and by considering interface effects of thermal environment on the energy spectrum of the finite chain.

At the heart of the Hill’s nanothermodynamics lie the consideration of a macroscopic ensemble of equivalent, identical, non-interacting small systems, whose addition or removal energy cost is associated with so-called subdivision potential~\cite{hillbook,doi:10.1002/ijch.196500008}. An alternative formulation of thermodynamics of a system, smaller relative to its surroundings, is based upon the idea of perturbation of the system spectrum by the thermal environment~\cite{Elcock_1957,Rushbrooke}. Canonical thermalization of the small system at the environment temperature can be properly explained~\cite{miguel1,miguel2,miguel_temperature-dependent_2015} by applying such a method of temperature dependent energy levels (TDELs). Hill’s nanothermodynamics and method of TDELs are connected to each other by recognizing the subdivision potential as the thermal perturbation by the environment~\cite{miguel1}. On the other hand, TDEL method predicts a different amount of heat exchange between the heat bath and the system, as it takes into account the contribution of the bath-system interface as another energy dissipation channel to the heat exchange~\cite{yamano2,yamano_effect_2017}, consistent with the extended second law of thermodynamics~\cite{Shental_2009}. Similar to the analysis of heat transfer efficiency in thermoelectric devices with TDELs~\cite{yamano_effect_2017}, we examine effects of system-bath interface in the operation of our topological nanothermodynamic heat engine.

Using the Hill's nanothermodynamic methods, we distinguish the bulk and boundary contributions in the Kitaev chain engine. It is found that when the whole system runs in an Otto engine cycle, bulk and boundary undergo their own non-Otto engine and refrigerator cycles, making positive and negative contributions to the overall work output, respectively. In addition, it is pointed out that three independent Otto cycles can also be introduced between the same hot and cold baths for the total system, bulk and boundary. We can have an Otto refrigerator separately at the boundary, while bulk and total system can be in their own independent Otto engine cycles. When we adopt the TDELs, the heat exchange between the baths and the chain is modified, consistent with the generalized second law. Both the input and output heat in the engine operation is reduced,   overall work output is also decreased, when the energy loss in the system-bath interface is taken into account. 

This paper is organized as follows.  We briefly review the Hill's nanothermodynamics and the framework of TDELs that we use in two subsections in Sec.~\ref{sec:methods}. Our model system, finite-length Kitaev Chain is presented in Sec.~\ref{sec:model}. The results and discussions are given in Sec.~\ref{sec:res} in four subsections. We conclude in Sec.~\ref{sec:conc}. 

\section{Methods}\label{sec:methods}

We summarize here the key points of two major historical approaches to extend thermodynamic description to finite size systems: (i)
Hill's nanothermodynamics~\cite{hillbook} and (ii) temperature-dependent energy levels~\cite{miguel1,yamano_effect_2017,yamano2}. We shall restrict ourselves using these two general methods to develop our results while we remark that there are other approaches to the thermodynamics of small systems which are beyond the scope of present paper~\cite{lebowitz_thermodynamic_1961,sisman,bedeaux_hills_2018}. 
\subsection{Hill's Nanothermodynamics}\label{sec:nanothermo}

In finite systems, extensivity of free energies is broken due to
boundary effects. 
Hill's idea is to consider an ensemble of equivalent, idential and non-interacting small systems, so that ordinary thermodynamics can be used by introducing so-called subdivision potential, $\epsilon$
to represent the energetic cost of adding another replica of the small system to the ensemble. 

Let us assume the ensemble is a grand canonical one, in an environment consisting of a particle and a heat bath characterized by temperature $T$ and
chemical potential $\mu$, respectively. The Gibbs differential relation of the ordinary thermodynamics can now be employed to write
\begin{equation}\label{eq:dEt}
	dE_t=TdS_t-pMdV+\mu dN_t+\epsilon dM,
\end{equation}
where $M$ is the number of replicas of the small system of volume $V$ and pressure $p$ in the ensemble. Total internal energy, entropy, and the number of the particles in the ensemble are denoted by $E_t,S_t$, and $N_t$, respectively. The $\epsilon$ in the last term could be interpreted as a ``system'' chemical potential or
as the work done by expanding the ensemble by adding another replica to it. For the latter, the ``integral pressure'', $\hat p$ is defined, by taking $\epsilon$ as the grand potential of a single system, with $\epsilon=-\hat p V$ 
by Hill~\cite{doi:10.1002/ijch.196500008}.

Integrating Eq.~(\ref{eq:dEt}) by the Euler theorem, for fixed $T,p,\mu,\epsilon$ then
dividing the result by $M$ we find 
\begin{equation}\label{eq:hill-E}
E=TS-p V+\mu N+X.
\end{equation}
Here $E$ and $N$ are the mean energy and mean number of particles for a single small system in the ensemble; while $S$ is the entropy which the same for each small system~\cite{hillbook,doi:10.1002/ijch.196500008}.
Due to the last term $X:=(p-\hat{p})V$, which is called as subdivision potential, $E$ is not a linear homogeneous function of
$S,V,N$. In general $\hat p$ is size dependent, but the leading order term is independent of $V$. The extensivity is broken, while the ordinary thermodynamics can be recovered in the limit $X\rightarrow 0$.
The corresponding differential Gibbs relation is then found to be
\begin{equation}\label{eq:hill-dE}
dE=TdS-pdV+\mu dN,
\end{equation}
with the generalized Gibbs-Duhem equation
\begin{equation}\label{eq:hill-gibbsDuhem}
dX=-SdT+Vdp-Nd\mu,
\end{equation}
while the differential relation for $\epsilon$ is the same as the
ordinary thermodynamic one,
\begin{equation}\label{eq:hill-deps}
d\epsilon=-SdT-pdV-Nd\mu.
\end{equation} 

Connection to statistical mechanics can be made by
\begin{equation}
\epsilon = - k_BT\ln \Xi,
\end{equation}
where 
\begin{equation}
\Xi(\mu,T,V)=\Tr\exp[-\beta(H-\mu N)],
\end{equation}
is the grand canonical partition function for a single small
system described by the Hamiltonian $H$ in the ensemble
with $\beta=1/k_BT$.

Equations (\ref{eq:hill-E})-(\ref{eq:hill-gibbsDuhem}) constitute the essential relations of Hill's nanothermodynamics. We remark two subtle points that first, it is assumed that the small systems can be thermalized to the bath temperature $T$ and Hill's thermodynamics does not give any mechanism for this thermal equilibration; and second, subdivision potential is a phenomoelogical term presented without any microscopic origin. These two issues have been addressed within the framework of temperature dependent energy levels method.
\subsection{Temperature-dependent energy levels}\label{sec:tdel}
Method of TDELs has been proposed as a fast and convenient way of statistical mechanical calculations for an assembly of systems~\cite{Rushbrooke,Elcock_1957}
and applied to semiconductors
~\cite{VARSHNI1967,PhysRevB.30.5766,Allen_1976,Patrick_2014}, superfluids~\cite{PhysRevLett.119.256802}, optomechanical oscillators~\cite{kolar_optomechanical_2017}, heat losses in thermoelectric systems~\cite{yamano2,yamano_effect_2017}, thermalization
of finite-size systems~\cite{miguel1,miguel2,miguel_temperature-dependent_2015}. TDELs can be associated with a temperature-dependent effective Hamiltonian, or so called ``Hamiltonian of mean force'' arising as the result of a prior averaging over certain possible
microstates of the assembly in thermal equilibrium~\cite{seifert_first_2016,talkner_open_2016}. For example, electronic energy
levels in semiconductors change with the temperature due to interaction of the electron with the lattice subject to thermal expansion~\cite{PhysRevB.30.5766}.

In the presence of TDELs, the energy transferred from the heat bath to the system cannot be completely identified as heat. As the system
temperature changes to match with that of the heat bath, the energy
gaps between the energy levels may change, in addition to their  populations. Writing the mean energy as $E=\langle H(T) \rangle
=\Tr (\rho H(T))$ and keeping all the other parameters constant, the infinitesimal mean energy change for a small system 
in contact with a heat bath then reads~\cite{yamano2,yamano_effect_2017,miguel1,miguel2,miguel_temperature-dependent_2015}, 
\begin{equation}\label{eq:TDEL_dU}
dE=TdS_T+\left\langle \frac{ \partial H}{\partial T}\right\rangle dT.
\end{equation} 
Here the first term is the heat associated with the population changes of the TDELs with the corresponding entropy $S_T$. When the energy gaps change with $T$, the second term cannot be interpreted as heat; it represents the work applied to change the energy gaps. Such a mechanism has been discussed in detail to explain thermalization of small systems~\cite{miguel1,miguel2,miguel_temperature-dependent_2015} and an effective heat describing the actual heat transfer into the small system from the thermal surroundings is defined by~\cite{yamano2,yamano_effect_2017,miguel1,miguel_temperature-dependent_2015}
\begin{equation}\label{eq:dQeff}
\delta Q_{\text eff}=\delta Q-\left\langle \frac{\partial H}{\partial T}\right\rangle dT.
\end{equation} 
This equation originates from the first law modified for a small system; rewriting it as $dS_T=dQ/T-\langle \partial H/\partial T\rangle dT/T$ one recovers the modified second law of information transfer channels~\cite{Shental_2009}. The entropy transferred from a heat source into the system through the boundary is balanced by the entropy lost at the boundary. The boundary acts as an energy channel interfacing the  heat source and the system. 

Hill's nanothermodynamics and the method of TDELs can be connected to each other by recognizing that subdivision potential
can be determined by~\cite{miguel1}
\begin{equation}
X=T\left\langle \frac{\partial H}{\partial T}\right\rangle.
\end{equation}
We can use the calculations of Hill's nanothermodynamics to evaluate the heat transfer as described by
the TDEL method according to
\begin{equation}\label{eq:Qeff}
Q_{\text eff}=\int T dS - \int\frac{X}{T}dT.
\end{equation} 
We emphasize that the first term is written in terms the $S$ of Hill's nanothermodynamics according to which we have $dE=dQ=TdS$. TDELs method yields a correction term associated with the work done on
the energy gaps such that some heat is dissipated at the system-bath interface and only an effective heat is received by the system.
 \section{Model: Kitaev Chain in Otto Cycle}\label{sec:model}
Kitaev chain is the one dimensional topological superconductor model~\cite{kitaev}, which consists of spinless fermions, described by the Hamiltonian
\begin{eqnarray}\label{eq:HKC}
	H=-\mu\sum_{i=1}^N a^\dagger_{i}a_{i} 
	-\sum_{i=1}^{N-1} \Big(t a^\dagger_{i} a_{i+1}-
	\Delta a_{i+1} a_{i} + h.c.\Big).
\end{eqnarray}
Here, $i=1...N$ labels the $N$ sites on the one-dimensional finite-length Kitaev chain. $\mu$ is the chemical potential, $t$ is the hopping parameter, and $\Delta$ is the superconducting pairing parameter. The fermionic annihilation (creation) operators, $a_{i}(a^\dagger_{i})$ satisfy the anti-commutation relations $\{a_{i},a^\dagger_{j}\}=\delta_{ij}$.
The Kitaev chain can be found in 
a trivial and a topological phase for 
$|\mu|<2t,$ and  $|\mu|>2t$, respectively; and a topological phase
transition happens at $|\mu|=2t$. The Kitaev chain hosts a pair of Majorana fermions when it is in the topological phase.

Following the Hill's nanothermodynamics an ensemble of 
identical, equivalent, non-interacting copies of a finite-length
Kitaev chain will be considered. While for the TDEL approach,
we consider a macroscopic Kitaev chain decomposed into two parts: One part is taken as the finite-length system, and the other is an interface lead which is sufficiently long enough to be assumed in thermal equilibrium with a heat bath~\cite{Elcock_1957}. TDEL arise as the
result of averaging over the interface microstates and describe
thermodynamically consistent heat exchange during the thermalization of the finite-length chain at the heat bath temperature.

Bulk and boundary thermodynamical properties of the finite-length Kitaev chain can be investigated using the Hill's nanothermodynamic framework. For that aim, the following ansatz for the grand potential is introduced~\cite{quelle}
\begin{equation}
\Phi(\mu,T,L)=\Phi_c(\mu,T)L+\Phi_0(\mu,T).
\label{ansatz1}
\end{equation}
Here $\Phi_cL=-pL$ is the bulk grand potential, which is extensive, while
$\Phi_0=X$ is the subdivision potential emerging due to the finite lenght of the chain. The corresponding entropies obey a similar relation,
\begin{equation}
S = S_c L+ S_0.
\label{eq:Sbulkedge}
\end{equation}
Here $S$ is the total entropy of the system, and  $S_cL$ and $S_0$ are bulk and boundary contributions, respectively. 

We first find the eigenvalues of the Hamiltonian in Eq.~(\ref{eq:HKC}) then evaluate the total entropy
$S$ of the chain. $S_c$ and $S_0$ are determined by using a linear fit to $S$ for an $n$-site chain (we take the unit length of the chain as $1$ so that $L=n$) in the interval $200<n<225$. The length is chosen to be sufficiently
large to make the linear fit a valid approximation~\cite{quelle}. 
Repeating the procedure for different temperatures $T-S$ relation is found.
The thermodynamic cycle we consider will be that of an Otto engine. Otto cycle consists of two isentropic (adiabatic) and two isochoric (isoparametric) stages.
We take the hopping parameter $t$ as the control parameter for
the Otto cycle. The $S-T$ curves at different $t$ will be used to determine the explicit cycle diagram. 
and the associated work output In the following subsections the corresponding work and efficiency of the Kitaev chain Otto cycle will be investigated by special attention to distinguish separate
bulk and boundary contributions and by considering the effect of bath-system interface as an energy channel
modifying the heat exchange between the heat baths and the Kitaev chain. 
\section{Results and Discussion}\label{sec:res}
In our calculations we consider a Kitaev chain of length $n=225$ with a superconducting pairing parameter $\Delta=0.25$. Chemical potential is fixed at $\mu=0.5$, and $t$ is used as the control parameter for which TPT takes place at $t=0.25$. The energy parameters are scaled by $ \kB/\hbar v=1,$ where $v$ is the velocity of the excitations~\cite{universality}.
\subsection{Work and efficiency of finite-length Kitaev chain Otto Engine} \label{sec:Total}
\begin{figure}[t!]
	\centering
	\includegraphics[width=8.3 cm]{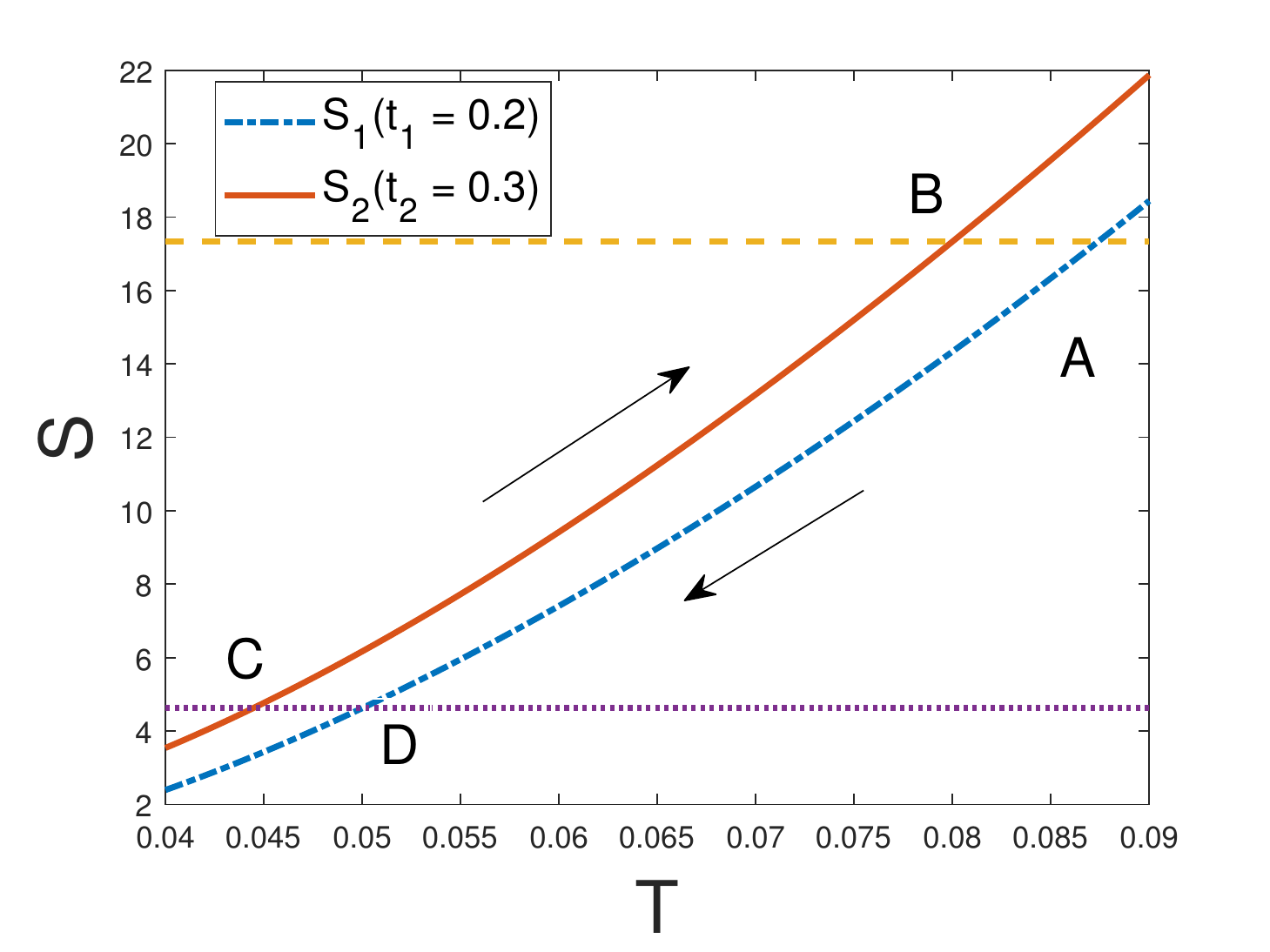}
	\caption{(Color Online) Entropy-temperature ($S-T$) curves of a finite-length Kitaev chain according to Hill's nanothermodynamics. The chain has $n=225$ sites and characterized with a superconducting pairing parameter $\Delta=0.25$. The chemical potential is fixed to $\mu=0.5$. The solid red curve is for the hopping  parameter $t_2=0.3$ while the blue dash-dotted one is for $t_1=0.2$. An  Otto cycle can be defined using the segments between the points $A,B,C,$ and $D$  determined by the intersection of the curves and the yellow dashed and purple dotted constant entropy  lines. The cycle operates between a hot bath at temperature $T_B = 0.08$ and a cold bath at temperature 
	$T_D = 0.05$ in the clockwise direction. The arrows indicate the direction of the heat engine cycle producing positive work.}
	\label{fig:ST1}
	\end{figure}

Let us first consider the finite-length Kitaev chain as a whole.
The $T-S$ curves of the chain at $t_1=0.2$ (blue dot-dashed) and $t_2=0.3$ (solid red) are plotted in Fig.~\ref{fig:ST1}. 
The horizontal lines are the constant entropies $S=S_2(T_B)=S_1(T_A)$
and $S=S_2(T_C)=S_1(T_D)$, where $T_x$ denote the temperature of the chain at the point $x=A,B,C,D$. The lines are fixed by the hot bath and cold bath temperatures, $T_B = 0.08$ and $T_D = 0.05$, respectively.  The Otto cycle is then determined by finding the 
intermediate temperatures, $T_A$ and $T_C$ from the constant entropy conditions.

The heat exchanges of the chain with the heat baths at the isoparametric stages $A\rightarrow D$ (cold isochore) and $C\rightarrow B$ (hot isochore) are calculated by
\begin{eqnarray}\label{eq:heat}
Q_{\text{in}}= \int_{T_C}^{T_B}T\frac{dS_2}{dT}dT,\quad
Q_{\text{out}}= \int_{T_A}^{T_D}T\frac{dS_1}{dT}dT,
\end{eqnarray} 
where the entropies of the chain for $t=t_1$ and $t=t_2$ are distinguished by $S_1$ and $S_2$, respectively.
The net work performed by the cycle is then calculated by
$W=Q_{\text{in}}+Q_{\text{out}}$. Positivity of $W$ is required
for heat engine operation while a negative work output would be
the case of refrigerator behavior.

\begin{figure}
	\centering
	\includegraphics[width=8.3 cm]{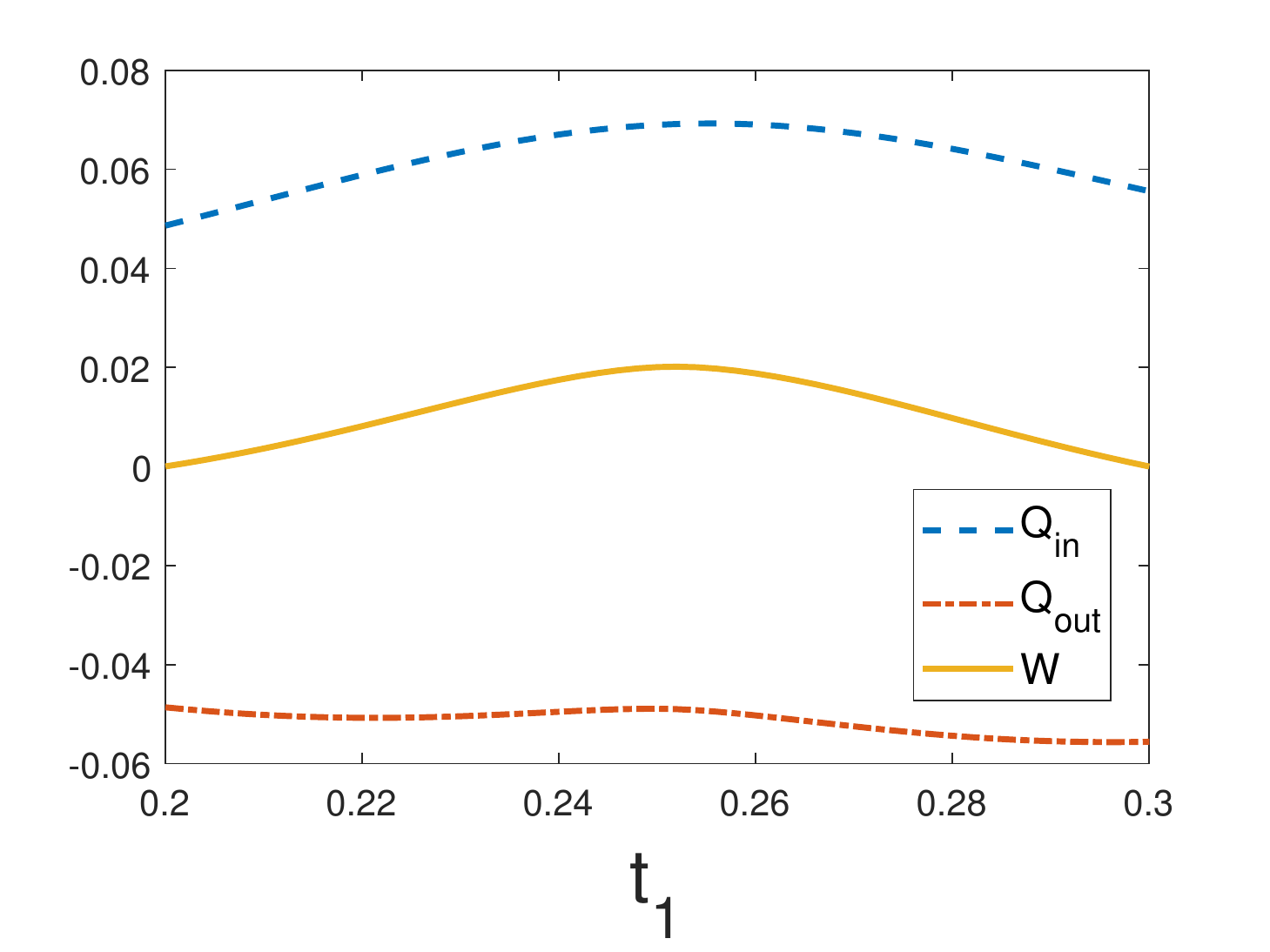}	
	\caption{(Color online)~Absorbed heat $Q_{\text{in}}$ (blue dashed), ejected heat $Q_{\text{out}}$ (red dotted dashed), and net work output $W$ (yellow solid) are given as a function of the hopping parameter $t_1$ of the cold isochore for an Otto cycle working between a hot bath at temperature $T_B= 0.08$ and a cold bath at temperature $T_D = 0.05$.}
	\label{fig:Total1}
\end{figure}

We now fix the parameter of the hot isochore $t_2=0.3$ and vary the one for the cold isochore $t_1\equiv t$ from $0.2$ to $0.3$.
Using the construction of the Otto cycle described in Fig.~\ref{fig:ST1}, we evaluate the work output and the efficiency of the cycle. The injected and ejected heat are plotted, together with the work output, in Fig.~\ref{fig:Total1} for the range of $t_1$. We observe that at $t = 0.25$, which is the critical point of
TPT, the heat absorbed by the Kitaev chain and the associated  work output of the cycle becomes maximum. For this range of $t_2$, $Q_{\text{in}}>0$ and $Q_{\text{out}}<0$ so that the cycle can be properly described as a heat engine operation. 

	
\begin{figure}
	\centering
	\includegraphics[width=8.3 cm]{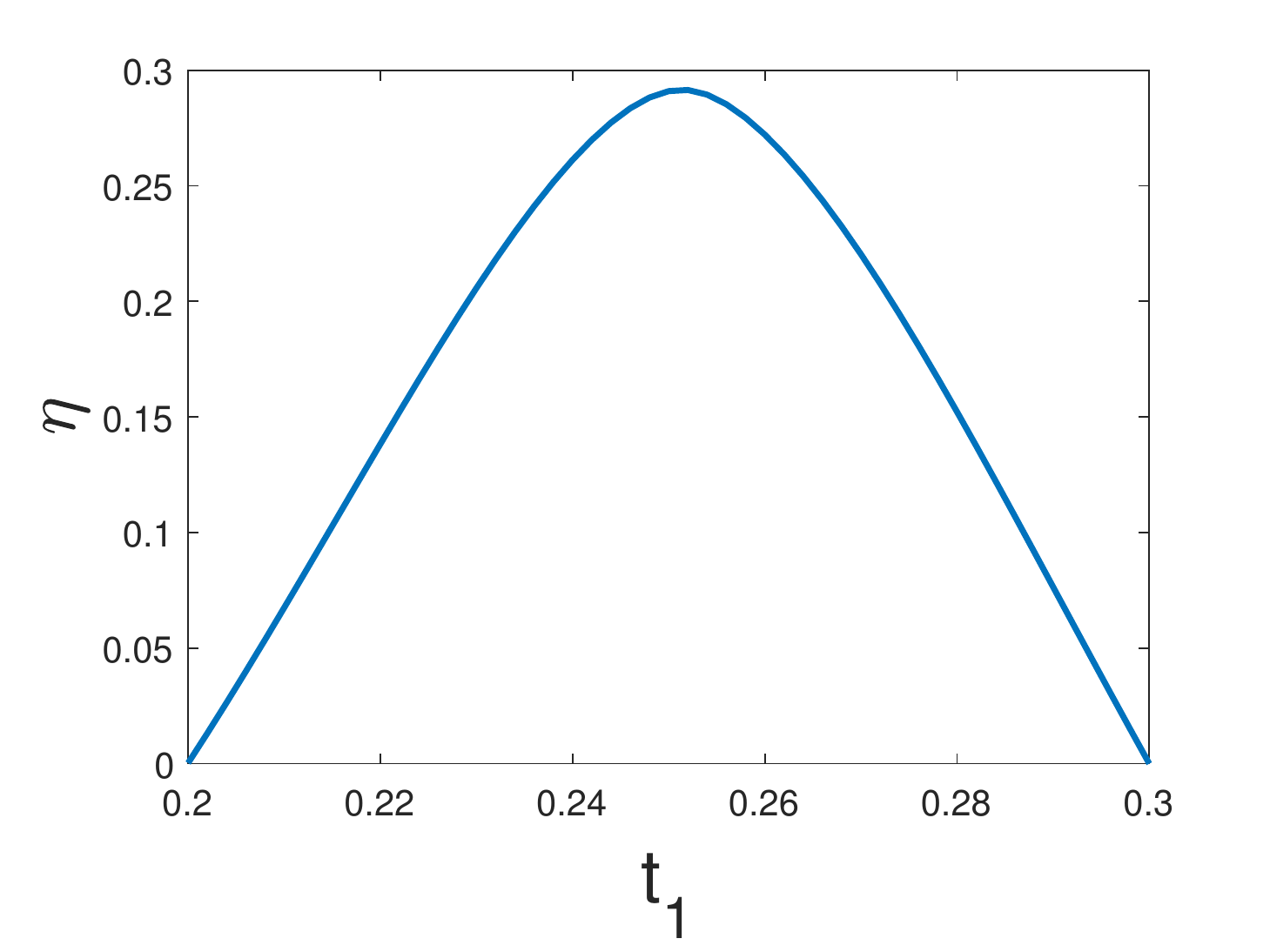}	
	\caption{(Color online)~Efficiency $\eta$ of finite Kitaev chain heat engine is given as a function of the hopping parameter $t_1$ of the cold isochore for an Otto cycle working between a hot bath at temperature $T_B= 0.08$ and a cold bath at temperature $T_D = 0.05$.}
	\label{fig:efftotal}
\end{figure}

The efficiency is plotted in Fig.~\ref{fig:efftotal} and is maximum at TPT. TPT enhances finite the Kitaev chain heat engine work output as well as its efficiency.

\subsection{Bulk and boundary contribution 
	to the work output of the Kitaev chain heat engine}\label{sec:bulkboundary}

Hill's nanothermodynamic framework allows us to examine the bulk and boundary contributions to the net work output of the Kitaev chain heat engine. For that aim, we use Eq.~(\ref{eq:Sbulkedge}) 
to write the total heat exchange
\begin{equation}
Q=\int TdS = \int TdS_c L+ \int TdS_0 = Q_{c}L+Q_{0}, 
\end{equation}
in terms of the bulk and boundary contributions, $Q_c L$ and $Q_0$, respectively.

\begin{figure*}[t!]
	\centering
	\subfloat[]{\includegraphics[width=8.3 cm]{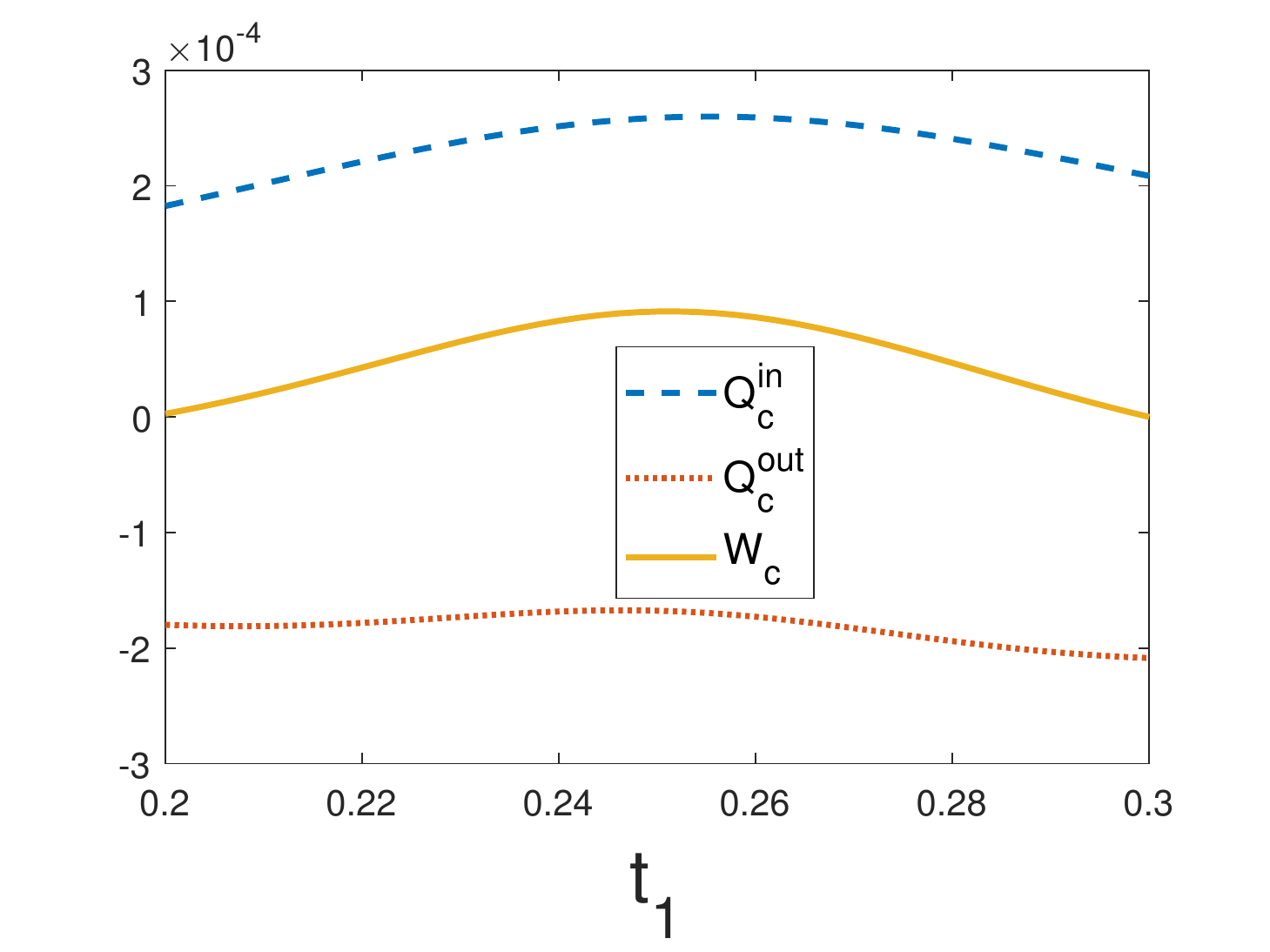}	\label{fig:bulk_heatwork1}}\qquad
	\subfloat[]{\includegraphics[width=8.3 cm]{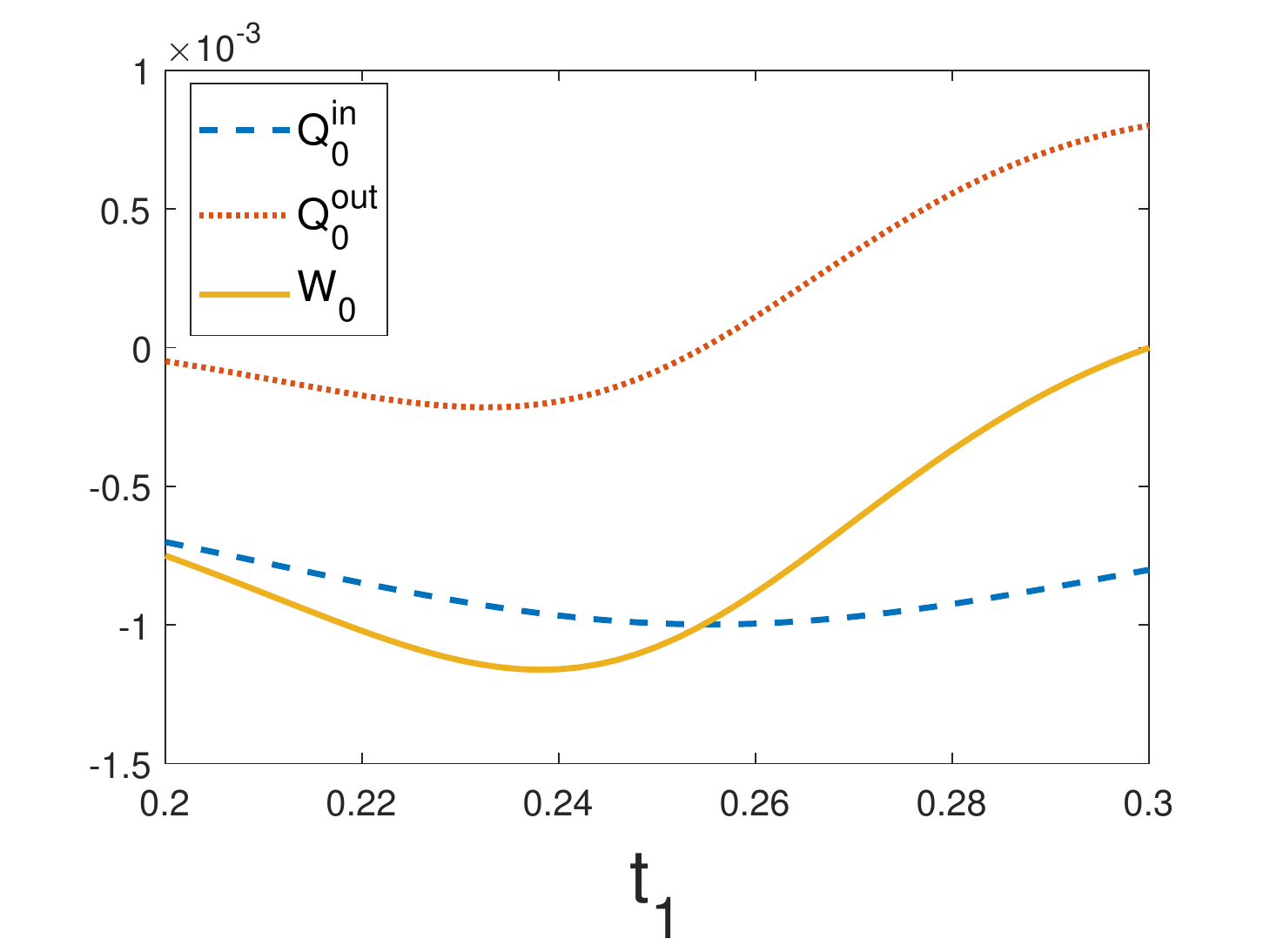}	\label{fig:edge_heatwork1}}
	\caption{(Color Online)~(a) The bulk and (b) the boundary contributions to the injected $Q^{\text{in}}_c$, $Q^{\text{in}}_0$ (blue dashed) and  ejected heat, $Q^{\text{out}}_c$, $Q^{\text{out}}_0$ (red dotted), and to the work output $W_c$, $W_0$ (yellow solid), respectively, as a function of the hopping parameter $t_1$ of the cold isochore of a finite-length Kitaev chain in an Otto cycle, operating between a hot bath at temperature $T_B= 0.08$ and a cold bath at temperature $T_D = 0.05$. }
	\label{fig:bulk_boundary_q}
\end{figure*}

For the same Otto cycle, described in Fig.~\ref{fig:ST1}, the bulk contributions to injected and ejected heat are plotted in Fig.~\ref{fig:bulk_heatwork1}. Behavior of the heat absorbed by the bulk depending on $t$ is qualitatively the same with that for the entire chain (cf.~Fig.~\ref{fig:Total1}).

While the dependence of injected and ejected heat on the hopping parameter $t_1$ of the cold isochore do not exhibit any qualitative difference for bulk and the entire chain, boundary contribution is remarkably different from the both. Fig.~\ref{fig:edge_heatwork1} plots the boundary contribution to the heat transfers. Boundary contribution is about an order of magnitude smaller than the bulk contribution, and hence cannot change the qualitative behavior dictated by the bulk contribution.
Boundary contribution to the work output is negative for the range of $t_1$ considered, in contrast to the bulk contribution, which is positive. Boundary reduces the injected heat in the topological phase but enhances it in the trivial phase. Ejected heat benefits from the boundary in both phases. Contributions of both the bulk and the boundary to the work output of the engine are maximum at the critical point of TPT. 

We remark that the bulk and boundary contributions to the work output of the entire finite-length Kitaev chain are not associated with their separate Otto engines per se. How to identify such independent operating bulk, boundary, and total Otto cycles will be the question we address in the next subsection.

\subsection{Three Otto cycles with a single finite-length Kitaev chain}\label{sec:threeOtto}

It is possible to identify three independently running Otto cycles, associated with the total, bulk and boundary separately. For that aim we examine the entropy-temperature curves, $S_c-T$, $S_0-T$, and $S-T$ separately. We consider the same temperature range for the cycles as in the previous discussions, with a hot bath at temperature $T_B=0.08$ and a cold bath at temperature at $T_D=0.05$. It is possible to define Otto cycles for the total and bulk sections of the engine in a wider range of parameter regimes, however in order to consider an Otto cycle for the boundary, a limited parameter range exists for the temperatures considered. Thus, we are limited to the parameter range $t_1=0.29$ and $t_2=0.3$ to be able to construct three Otto cycles making use of total system, its bulk and boundary independently. We note that this parameter range is where the Kitaev chain is in its normal phase. 

Accordingly, the Otto cycles can be determined for bulk and boundary, similar to the case of the total system. The cycles will be different due to the differences in the intermediate temperatures determined by isentropy conditions for bulk and boundary entropies $S_c,S_0$ and entropy of the total chain $S$ and hence would yield different work outputs. In general, our numerical investigations suggest that bulk and total system temperatures are close to each other while the boundary can have significantly different temperatures. Moreover, the entropy of the boundary can be negative for some parameter regimes~\cite{universality}. Through numerical investigations we find a narrow regime in the trivial phase, between $t_1=0.29$ and $t_2=0.3$ for which boundary can receive work through a refrigerator. 

\begin{figure*}[t!]
	\centering
	\subfloat[]{\includegraphics[width=8.3cm]{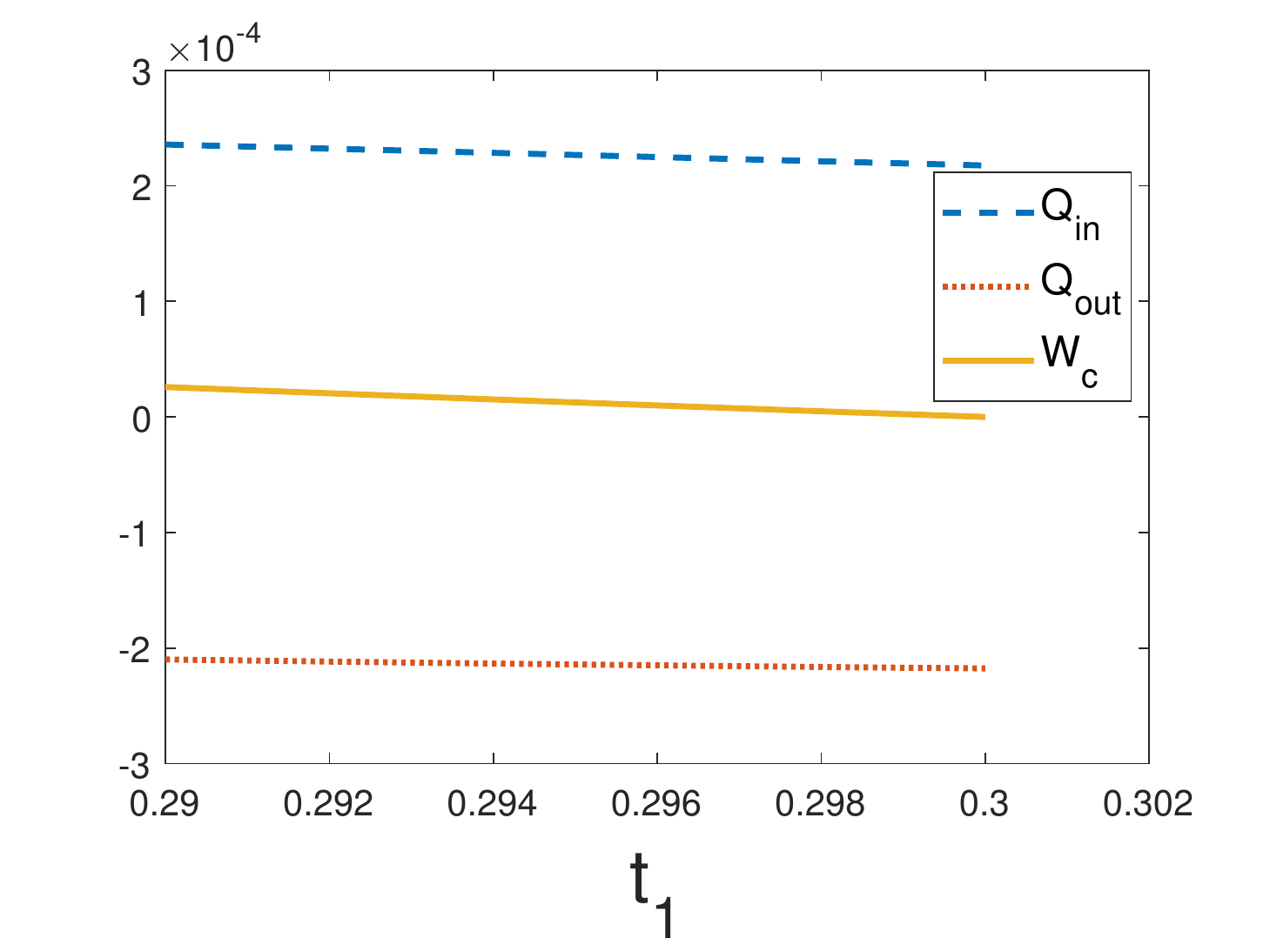}	    \label{fig:3Ottobulk}}\qquad
	\subfloat[]{\includegraphics[width=8.3cm]{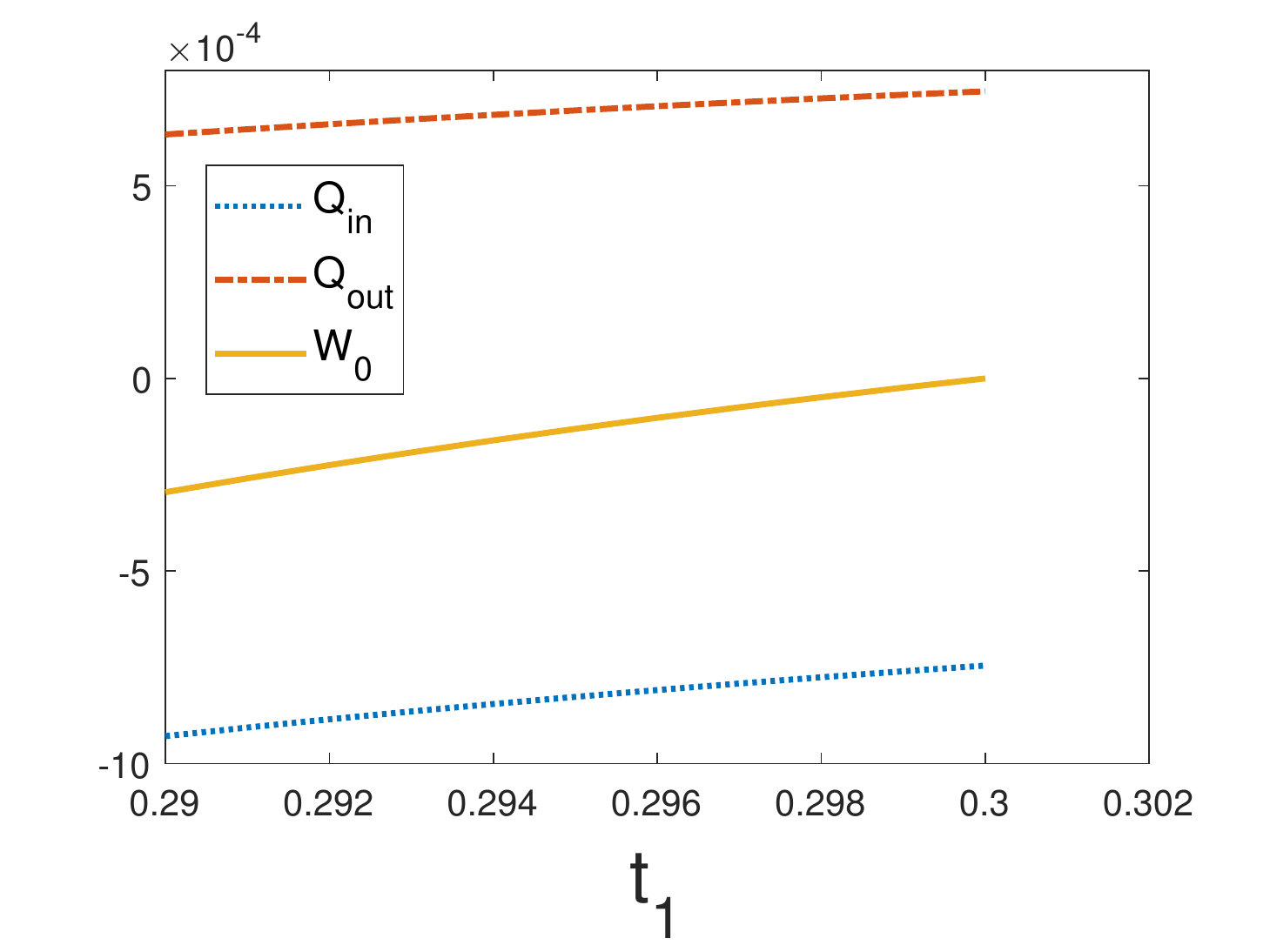}	\label{fig:3Ottoboundary}}
	\caption{(Color online)~ The injected $Q^{\text{in}}_c$, $Q^{\text{in}}_0$ (red dashed) and  ejected heat,$Q^{\text{out}}_c$, $Q^{\text{out}}_0$ (blue dotted) and  the work output $W_c$, $W_0$ (yellow solid) coressponding to (a) the bulk heat engine and (b) the boundary refrigerator, respectively, as a function of the hopping parameter $t_1$ of the cold isochore of a finite-length Kitaev chain in independent Otto cycles, operating between a hot bath at temperature $T_B= 0.08$ and a cold bath at temperature $T_D = 0.05$ in the parameter range between $t_1=0.29$ and $t_2=0.3$. Note that the intermediate temperatures for each engine would be different due to different entropy values leading to different isentropy conditions.}
	\label{fig:3Ottowork}
\end{figure*}

The work outputs are shown in Figs.~\ref{fig:3Ottobulk} and \ref{fig:3Ottoboundary}. For the temperature and parameter range considered, the bulk behaves like a heat engine for the Otto cycle, while the boundary becomes a refrigerator. For the clockwise direction of the cycle, the bulk receives heat as its temperature increases isoparametricaly from C to B, and releases heat going from point A to D. For the boundary, again for the clockwise direction of the cycle, the situation is reversed such that the boundary releases heat going from point C to B as its temperature decreases and vice versa from point A to D.

\begin{figure}[t!]
	\centering
	\includegraphics[width=8.3 cm]{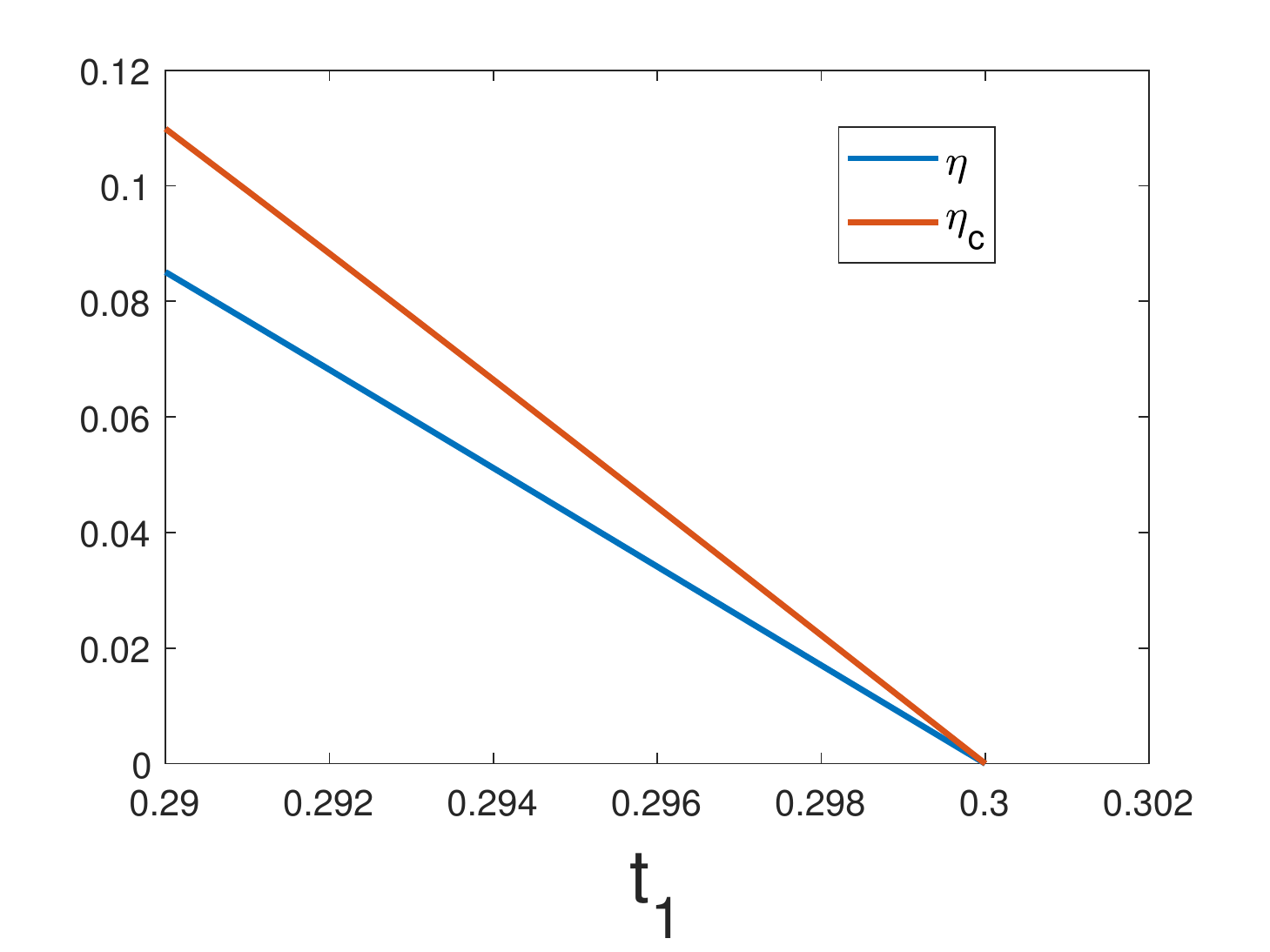}	
	\caption{(Color online)~Efficiencies of (a)~ a total system Otto heat engine (blue solid),  (b)~  a bulk Otto heat engine (red solid), $\eta,\eta_c$ respectively  are given as a function of the hopping parameter $t_1$ of the cold isochore for Otto cycles working between a hot bath at temperature $T_B= 0.08$ and a cold bath at temperature $T_D = 0.05$.}
	\label{fig:3Ottoeff}
\end{figure}

When we compare the efficiencies of the heat engine for the total system and its bulk, as in Fig.~\ref{fig:3Ottoeff}, we observe that the bulk heat engine is more efficient than the total. 

\subsection{Effective Work}\label{sec:effectivework}

Both Hill's approach and TDEL method are used to describe the thermodynamics of small systems. The connection between these two nanothermodynamic approaches is sustained by recognizing that the subdivision potential acts as the thermal perturbation by the environment. However, these two methods present different amount of heat exchange between a bath and system. 
Hill's nanothermodynamics allows for the contribution of bulk and boundary to be identified separately so that the thermodynamic properties of a finite system can be investigated for bulk, boundary, and the total system separately. We elaborate in Sec.~\ref{sec:methods} how three independent Otto cycles can be formed taking advantage of this identification. 

TDEL method, on the other hand, describes the boundary as an energy channel interfacing the bath and the system. As is eloborated in Sec.~\ref{sec:tdel}, the subdivision potential of Hill's nanothermodynamics can be used to calculate the heat dissipated through this boundary energy channel. This is with the aid of making the association $X=\phi_0$ in Eq.~(\ref{eq:Qeff})~\cite{miguel1}. 

Incoming and outgoing heats read
\begin{eqnarray}
Q_{\text{out(eff)}}= \int_{S_A}^{S_D}TdS-\int_{T_A}^{T_D}\frac{\Phi_0}{T}dT,
\label{eq:qouteff}
\end{eqnarray} 

\begin{eqnarray}
Q_{\text{in(eff)}}= \int_{S_C}^{S_B}TdS-\int_{T_C}^{T_B}\frac{\Phi_0}{T}dT.
\label{eq:qineff}
\end{eqnarray}

\begin{figure*}[t!]
	\centering
	\subfloat[]{\includegraphics[width=8.3cm]{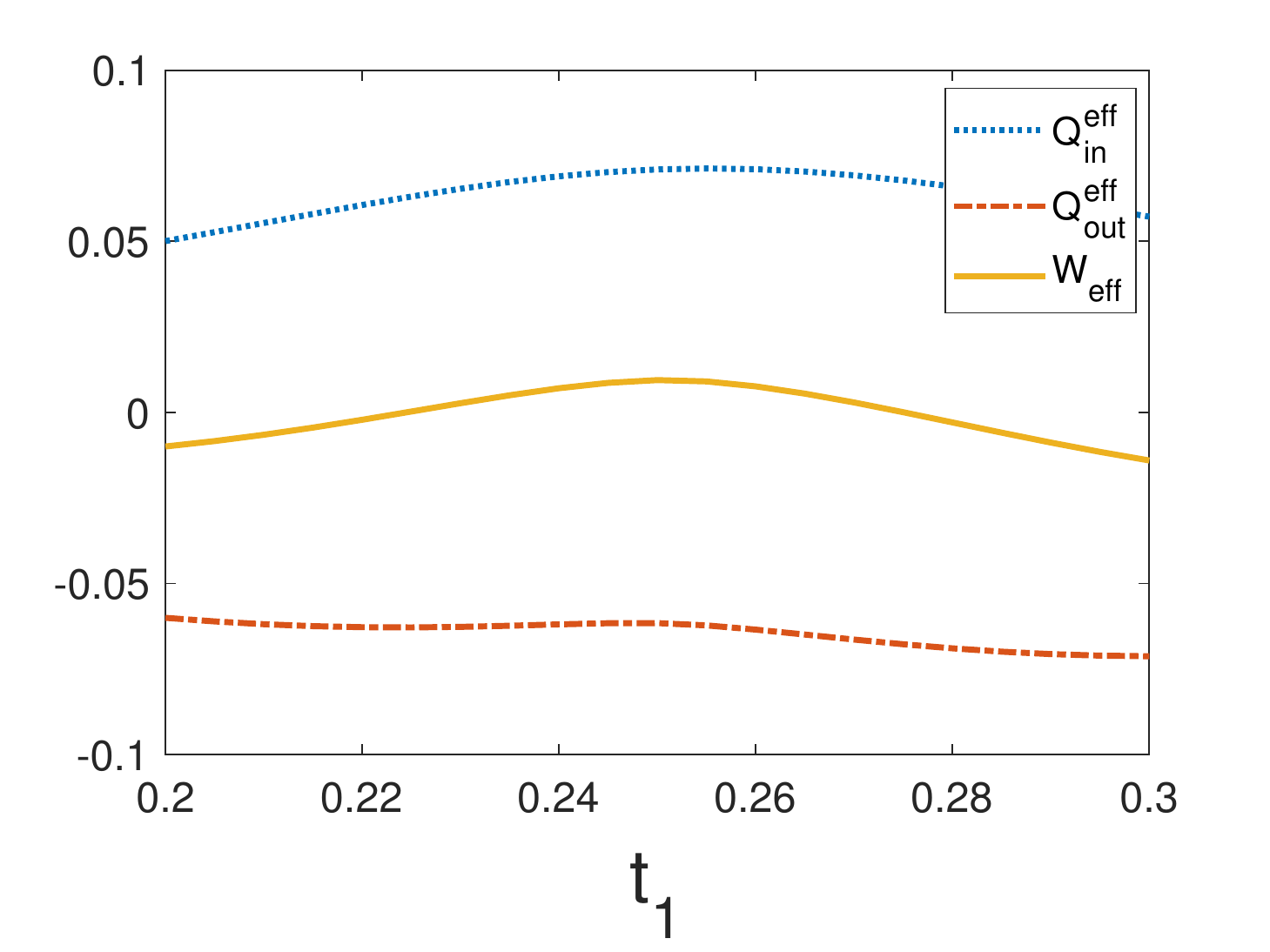}	    \label{fig:figure7a}}\qquad
	\subfloat[]{\includegraphics[width=8.3cm]{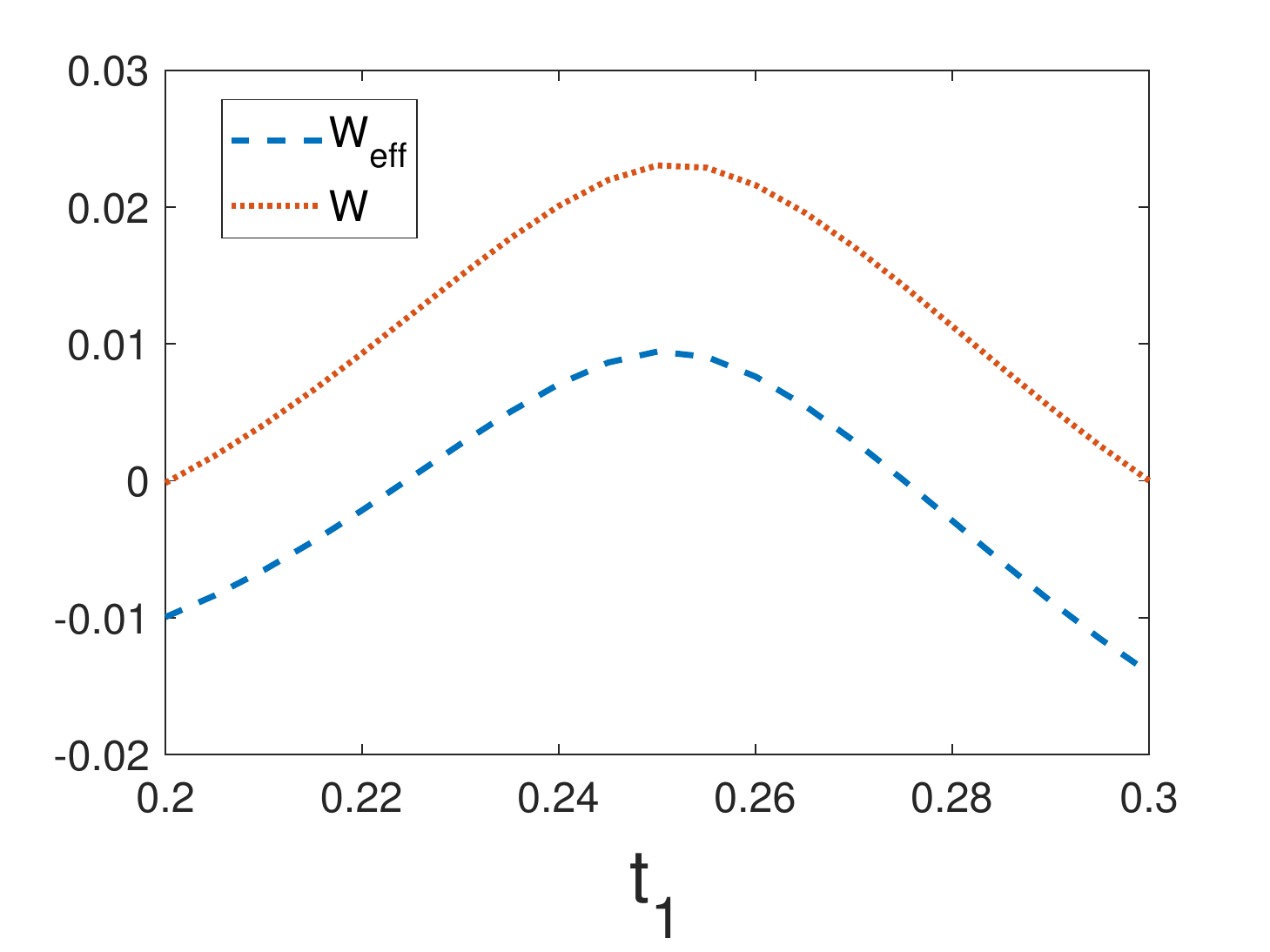}	    \label{fig:figure7b}}
	\caption{(Color online)~ (a)~Effective incoming heat $Q_{\text{in(eff)}}$ (blue-dotted), effective outgoing heat $Q_{\text{out(eff)}}$ (red dot-dashed) and effective work $W_{\text{eff}}$ (yellow solid), and(b) total net work  $W$ (red dashed) and effective work $W_{\text{eff}}$ (blue dotted) are given as a function of $t_1$ for an Otto cycle working between a hot bath at temperature $T_B= 0.08$ and a cold bath at temperature $T_D= 0.05$.}
	\label{fig:figure7}
\end{figure*}

The total work is
\begin{eqnarray}
W_{\text{eff}}=Q_{\text{in(eff)}}+Q_{\text{out(eff)}}.
\label{eq:workeff}
\end{eqnarray} 
 We calculate the injected and ejected effective heat and also the effective work for the Otto cycle in Fig.~\ref{fig:ST1} using Eqs.~(\ref{eq:qouteff})-(\ref{eq:workeff}). The results are plotted in Fig.~\ref{fig:figure7a}.

\begin{figure*}[t!]
	\centering
	\subfloat[]{\includegraphics[width=8.3cm]{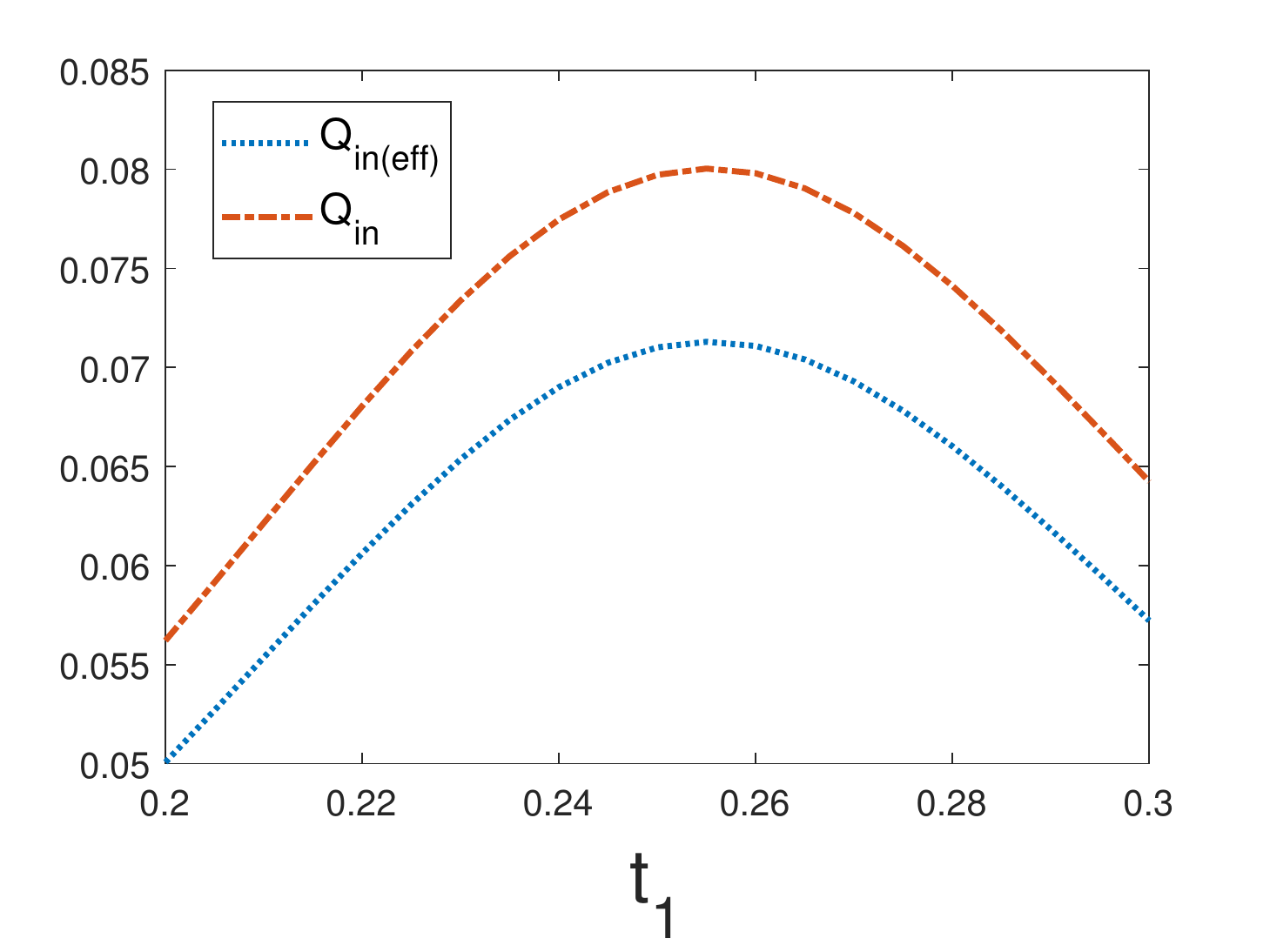}	    \label{fig:figure8a}}\qquad
	\subfloat[]{\includegraphics[width=8.3cm]{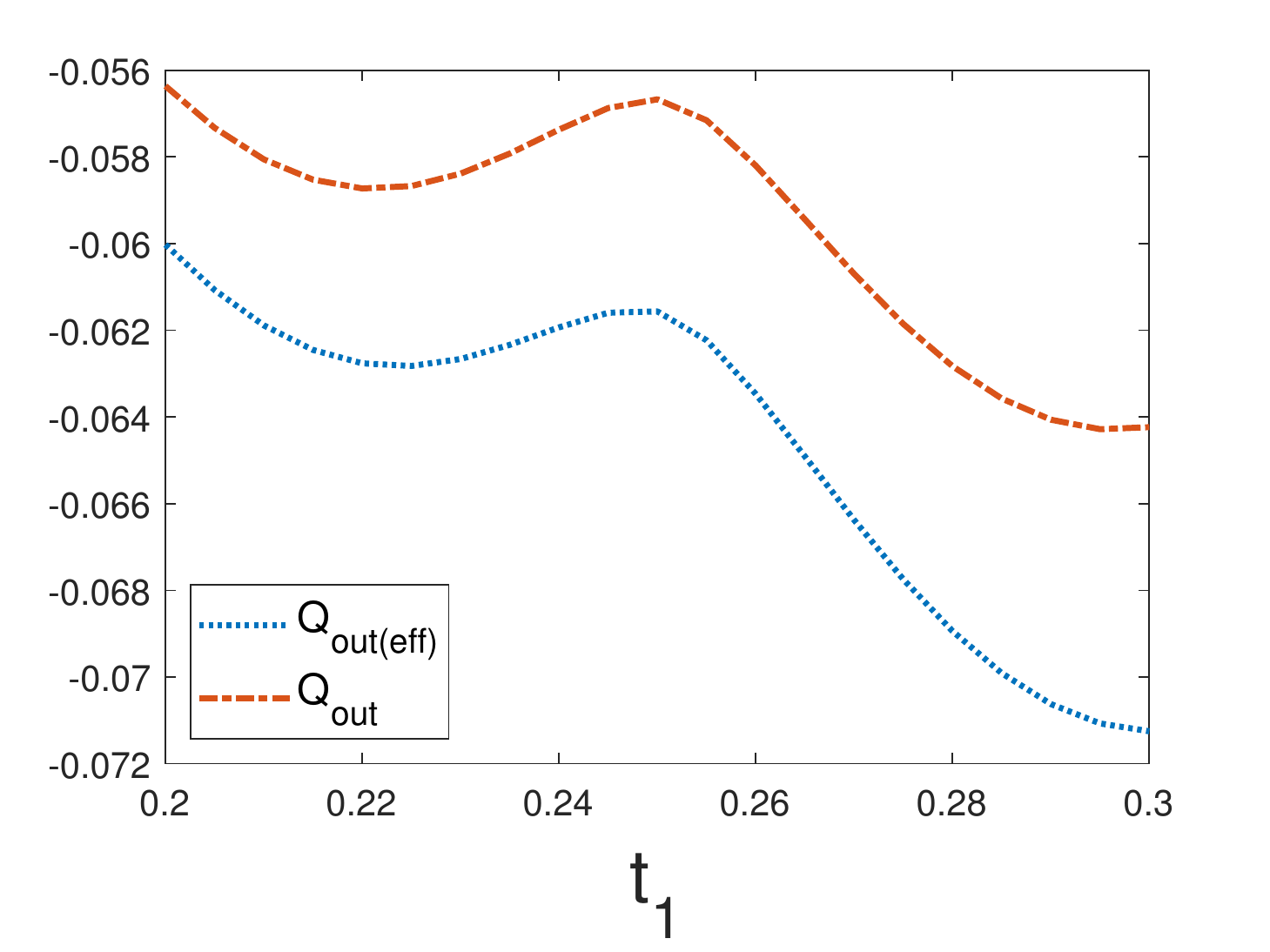}	    \label{fig:figure8b}}
	\caption{(Color online)~ (a) Absorbed heat $Q_{\text{in}}$ (red dot-dashed) and effective absorbed heat $Q_{\text{in(eff)}}$ (blue dotted), and (b)~ ejected heat $Q_{\text{out}}$ (red dot-dashed) and effective ejected heat $Q_{\text{out(eff)}}$ (blue-dotted)
		are given as a function of $t_1$ for an Otto cycle working between a hot bath at temperature $T_B= 0.08$ and a cold bath at temperature $T_D = 0.05$.}
	\label{fig:figure8}
\end{figure*}

 There is a signature of TPT in both $Q_{\text{in(eff)}}$ and $Q_{\text{out(eff)}}$ and in the effective work $W_{\text{eff}}$ as well at $t=\mu/2$ as a maximum point. At the critical point where the topological phase transition  takes place, the system absorbs the maximum amount of heat and the associated work gets maximum. We also present a comparison between work, calculated using Hill's nanothermodynamics, and effective work in Fig.~\ref{fig:figure7b}. Both work and effective work have qualitatively the same behavior, but the effective work  $W_{\text{eff}}$ acquires lower values for the same parameter range. Moreover, it is negative for $t\textless0.225$ and $t\textgreater0.28$ and operates as a refrigerator.
 The behavior of exchanged heats yielded by Hill's nanothermodynamics are compared with the effective heats in Fig.~\ref{fig:figure8}.
It is observed that both $Q_{\text{in(eff)}}$ and $Q_{\text{out(eff)}}$ are smaller than what we obtained by using Hill's nanothermodynamic formalism. However, they are similar qualitatively and show TPT sign at $t=\mu/2$.



\section{conclusions}\label{sec:conc}
We propose a heat engine which uses the finite Kitaev chain as its working substance. The heat engine is based on an Otto cycle where there are two isoentropic (adiabatic) and two isoparametric stages. The external parameter in the engine is the hopping parameter of the Kitaev chain.

We report that the topological phase transition of the finite Kitaev chain enhances both the total work output and the efficiency. We further investigate the thermodynamic properties of the finite-length Kitaev chain within two thermodynamic frameworks for finite size systems: Hill's nanothermodynamics and TDELs scheme. Based on Hill's nanothermodynamics, we identify the separate qualitative and quantitative features arising from the bulk and boundary in the work output of the Otto cycle. We found that bulk and boundary undergoes non-Otto cycles; bulk produces positive work, as the total system, while the boundary makes negative contribution to the total work and operates in a refrigerator cycle. 

We also point out the possibility of three Otto cycles working independently between two baths and making use of the total, bulk and boundary separately. This scheme only works in the normal phase of the Kitaev chain. While the total system and the bulk operate as heat engines, the boundary operates as a refrigerator. The bulk heat engine is more efficient than the total system heat engine.

In addition, we have calculated the effective heat exchanges and the effective work within the TDELs scheme. We find that the effective work is qualitatively lower than what is calculated by Hill's nanothermodynamics. Both the incoming and outgoing heat exchanges are reduced as well due to the energy dissipation in the system-bath interface.

 While Hill's nanothermodynamics allows for the identification of the bulk and boundary contributions in the heat engine operation clearly, TDELs provides proper assignment of heat exchanges between the heat baths and the finite system. We conclude that these two approaches compliment each other and can be used in conjunction for detailed modeling of thermal machines and experiments with topological finite systems.

\section*{Acknowledgements}
We acknowledge support by the Scientific and Technological Research Council of Turkey (T\"{U}B{\.I}TAK), 
Grant No.~ (117F097) and by the EU-COST Action (CA16221). E.Y. would like to thank Asl{\i} Tuncer for invaluable discussions.

\newpage


\begin{thebibliography}{35}%
	\makeatletter
	\providecommand \@ifxundefined [1]{%
		\@ifx{#1\undefined}
	}%
	\providecommand \@ifnum [1]{%
		\ifnum #1\expandafter \@firstoftwo
		\else \expandafter \@secondoftwo
		\fi
	}%
	\providecommand \@ifx [1]{%
		\ifx #1\expandafter \@firstoftwo
		\else \expandafter \@secondoftwo
		\fi
	}%
	\providecommand \natexlab [1]{#1}%
	\providecommand \enquote  [1]{``#1''}%
	\providecommand \bibnamefont  [1]{#1}%
	\providecommand \bibfnamefont [1]{#1}%
	\providecommand \citenamefont [1]{#1}%
	\providecommand \href@noop [0]{\@secondoftwo}%
	\providecommand \href [0]{\begingroup \@sanitize@url \@href}%
	\providecommand \@href[1]{\@@startlink{#1}\@@href}%
	\providecommand \@@href[1]{\endgroup#1\@@endlink}%
	\providecommand \@sanitize@url [0]{\catcode `\\12\catcode `\$12\catcode
		`\&12\catcode `\#12\catcode `\^12\catcode `\_12\catcode `\%12\relax}%
	\providecommand \@@startlink[1]{}%
	\providecommand \@@endlink[0]{}%
	\providecommand \url  [0]{\begingroup\@sanitize@url \@url }%
	\providecommand \@url [1]{\endgroup\@href {#1}{\urlprefix }}%
	\providecommand \urlprefix  [0]{URL }%
	\providecommand \Eprint [0]{\href }%
	\providecommand \doibase [0]{http://dx.doi.org/}%
	\providecommand \selectlanguage [0]{\@gobble}%
	\providecommand \bibinfo  [0]{\@secondoftwo}%
	\providecommand \bibfield  [0]{\@secondoftwo}%
	\providecommand \translation [1]{[#1]}%
	\providecommand \BibitemOpen [0]{}%
	\providecommand \bibitemStop [0]{}%
	\providecommand \bibitemNoStop [0]{.\EOS\space}%
	\providecommand \EOS [0]{\spacefactor3000\relax}%
	\providecommand \BibitemShut  [1]{\csname bibitem#1\endcsname}%
	\let\auto@bib@innerbib\@empty
	\bibitem [{\citenamefont {Hill}(1964)}]{hillbook}%
	\BibitemOpen
	\bibfield  {author} {\bibinfo {author} {\bibfnamefont {Terrell~L.}\
			\bibnamefont {Hill}},\ }\href@noop {} {\emph {\bibinfo {title}
			{Thermodynamics of {{Small Systems}}}}}\ (\bibinfo  {publisher} {{Courier
			Corporation}},\ \bibinfo {year} {1964})\BibitemShut {NoStop}%
	\bibitem [{\citenamefont {Hill}(1965)}]{doi:10.1002/ijch.196500008}%
	\BibitemOpen
	\bibfield  {author} {\bibinfo {author} {\bibfnamefont {Terrell~L.}\
			\bibnamefont {Hill}},\ }\bibfield  {title} {\enquote {\bibinfo {title}
			{Thermodynamics of small systems, part i},}\ }\href {\doibase
		10.1002/ijch.196500008} {\bibfield  {journal} {\bibinfo  {journal} {Isr. J.
				Chem.}\ }\textbf {\bibinfo {volume} {3}},\ \bibinfo {pages} {39--39}
		(\bibinfo {year} {1965})}\BibitemShut {NoStop}%
	\bibitem [{\citenamefont {Chamberlin}(2015)}]{e17010052}%
	\BibitemOpen
	\bibfield  {author} {\bibinfo {author} {\bibfnamefont {R.~V.}\ \bibnamefont
			{Chamberlin}},\ }\bibfield  {title} {\enquote {\bibinfo {title} {The big
				world of nanothermodynamics},}\ }\href {\doibase 10.3390/e17010052}
	{\bibfield  {journal} {\bibinfo  {journal} {Entropy}\ }\textbf {\bibinfo
			{volume} {17}},\ \bibinfo {pages} {52--73} (\bibinfo {year}
		{2015})}\BibitemShut {NoStop}%
	\bibitem [{\citenamefont {Hill}(2001{\natexlab{a}})}]{hill_perspective_2001}%
	\BibitemOpen
	\bibfield  {author} {\bibinfo {author} {\bibfnamefont {Terrell~L.}\
			\bibnamefont {Hill}},\ }\bibfield  {title} {\enquote {\bibinfo {title}
			{Perspective : Nanothermodynamics},}\ }\href {\doibase 10.1021/nl010010d}
	{\bibfield  {journal} {\bibinfo  {journal} {Nano Lett.}\ }\textbf {\bibinfo
			{volume} {1}},\ \bibinfo {pages} {111--112} (\bibinfo {year}
		{2001}{\natexlab{a}})}\BibitemShut {NoStop}%
	\bibitem [{\citenamefont {Hill}(2001{\natexlab{b}})}]{doi:10.1021/nl010027w}%
	\BibitemOpen
	\bibfield  {author} {\bibinfo {author} {\bibfnamefont {Terrell~L.}\
			\bibnamefont {Hill}},\ }\bibfield  {title} {\enquote {\bibinfo {title} {A
				different approach to nanothermodynamics},}\ }\href {\doibase
		10.1021/nl010027w} {\bibfield  {journal} {\bibinfo  {journal} {Nano Lett.}\
		}\textbf {\bibinfo {volume} {1}},\ \bibinfo {pages} {273--275} (\bibinfo
		{year} {2001}{\natexlab{b}})}\BibitemShut {NoStop}%
	\bibitem [{\citenamefont {Quelle}\ \emph {et~al.}(2016)\citenamefont {Quelle},
		\citenamefont {Cobanera},\ and\ \citenamefont {Smith}}]{quelle}%
	\BibitemOpen
	\bibfield  {author} {\bibinfo {author} {\bibfnamefont {A.}~\bibnamefont
			{Quelle}}, \bibinfo {author} {\bibfnamefont {E.}~\bibnamefont {Cobanera}}, \
		and\ \bibinfo {author} {\bibfnamefont {C.~Morais}\ \bibnamefont {Smith}},\
	}\bibfield  {title} {\enquote {\bibinfo {title} {Thermodynamic signatures of
				edge states in topological insulators},}\ }\href {\doibase
		10.1103/PhysRevB.94.075133} {\bibfield  {journal} {\bibinfo  {journal} {Phys.
				Rev. B}\ }\textbf {\bibinfo {volume} {94}},\ \bibinfo {pages} {075133}
		(\bibinfo {year} {2016})}\BibitemShut {NoStop}%
	\bibitem [{\citenamefont {Kempkes}\ \emph {et~al.}(2016)\citenamefont
		{Kempkes}, \citenamefont {Quelle},\ and\ \citenamefont
		{Smith}}]{universality}%
	\BibitemOpen
	\bibfield  {author} {\bibinfo {author} {\bibfnamefont {S.~N.}\ \bibnamefont
			{Kempkes}}, \bibinfo {author} {\bibfnamefont {A.}~\bibnamefont {Quelle}}, \
		and\ \bibinfo {author} {\bibfnamefont {C.~Morais}\ \bibnamefont {Smith}},\
	}\bibfield  {title} {\enquote {\bibinfo {title} {Universalities of
				thermodynamic signatures in topological phases},}\ }\href {\doibase
		10.1038/srep38530} {\bibfield  {journal} {\bibinfo  {journal} {Sci. Rep.}\
		}\textbf {\bibinfo {volume} {6}},\ \bibinfo {pages} {38530} (\bibinfo {year}
		{2016})}\BibitemShut {NoStop}%
	\bibitem [{\citenamefont {Viyuela}\ \emph
		{et~al.}(2014{\natexlab{a}})\citenamefont {Viyuela}, \citenamefont {Rivas},\
		and\ \citenamefont {Martin-Delgado}}]{delgado1}%
	\BibitemOpen
	\bibfield  {author} {\bibinfo {author} {\bibfnamefont {O.}~\bibnamefont
			{Viyuela}}, \bibinfo {author} {\bibfnamefont {A.}~\bibnamefont {Rivas}}, \
		and\ \bibinfo {author} {\bibfnamefont {M.~A.}\ \bibnamefont
			{Martin-Delgado}},\ }\bibfield  {title} {\enquote {\bibinfo {title}
			{Two-dimensional density-matrix topological fermionic phases: Topological
				uhlmann numbers},}\ }\href {\doibase 10.1103/PhysRevLett.113.076408}
	{\bibfield  {journal} {\bibinfo  {journal} {Phys. Rev. Lett.}\ }\textbf
		{\bibinfo {volume} {113}},\ \bibinfo {pages} {076408} (\bibinfo {year}
		{2014}{\natexlab{a}})}\BibitemShut {NoStop}%
	\bibitem [{\citenamefont {Viyuela}\ \emph
		{et~al.}(2014{\natexlab{b}})\citenamefont {Viyuela}, \citenamefont {Rivas},\
		and\ \citenamefont {Martin-Delgado}}]{delgado2}%
	\BibitemOpen
	\bibfield  {author} {\bibinfo {author} {\bibfnamefont {O.}~\bibnamefont
			{Viyuela}}, \bibinfo {author} {\bibfnamefont {A.}~\bibnamefont {Rivas}}, \
		and\ \bibinfo {author} {\bibfnamefont {M.~A.}\ \bibnamefont
			{Martin-Delgado}},\ }\bibfield  {title} {\enquote {\bibinfo {title} {Uhlmann
				phase as a topological measure for one-dimensional fermion systems},}\ }\href
	{\doibase 10.1103/PhysRevLett.112.130401} {\bibfield  {journal} {\bibinfo
			{journal} {Phys. Rev. Lett.}\ }\textbf {\bibinfo {volume} {112}},\ \bibinfo
		{pages} {130401} (\bibinfo {year} {2014}{\natexlab{b}})}\BibitemShut
	{NoStop}%
	\bibitem [{\citenamefont {Broeke}\ \emph {et~al.}(2018)\citenamefont {Broeke},
		\citenamefont {Kempkes}, \citenamefont {Quelle}, \citenamefont {Wang},
		\citenamefont {Paglione},\ and\ \citenamefont
		{Smith}}]{broeke_thermodynamic_2018}%
	\BibitemOpen
	\bibfield  {author} {\bibinfo {author} {\bibfnamefont {J.~J. van~den}\
			\bibnamefont {Broeke}}, \bibinfo {author} {\bibfnamefont {S.~N.}\
			\bibnamefont {Kempkes}}, \bibinfo {author} {\bibfnamefont {A.}~\bibnamefont
			{Quelle}}, \bibinfo {author} {\bibfnamefont {X.~F.}\ \bibnamefont {Wang}},
		\bibinfo {author} {\bibfnamefont {J.}~\bibnamefont {Paglione}}, \ and\
		\bibinfo {author} {\bibfnamefont {C.~Morais}\ \bibnamefont {Smith}},\
	}\bibfield  {title} {\enquote {\bibinfo {title} {Thermodynamic study of
				topological {Kondo} insulators},}\ }\href {http://arxiv.org/abs/1803.03553}
	{\bibfield  {journal} {\bibinfo  {journal} {arXiv:1803.03553}\ } (\bibinfo
		{year} {2018})}\BibitemShut {NoStop}%
	\bibitem [{\citenamefont {Fadaie}\ \emph {et~al.}(2018)\citenamefont {Fadaie},
		\citenamefont {Yunt},\ and\ \citenamefont {M\"ustecapl\ifmmode \imath \else
			\i \fi{}o\ifmmode~\breve{g}\else \u{g}\fi{}lu}}]{PhysRevE.98.052124}%
	\BibitemOpen
	\bibfield  {author} {\bibinfo {author} {\bibfnamefont {M.}~\bibnamefont
			{Fadaie}}, \bibinfo {author} {\bibfnamefont {E.}~\bibnamefont {Yunt}}, \ and\
		\bibinfo {author} {\bibfnamefont {\"O.~E.}\ \bibnamefont {M\"ustecapl\ifmmode
				\imath \else \i \fi{}o\ifmmode~\breve{g}\else \u{g}\fi{}lu}},\ }\bibfield
	{title} {\enquote {\bibinfo {title} {Topological phase transition in
				quantum-heat-engine cycles},}\ }\href {\doibase 10.1103/PhysRevE.98.052124}
	{\bibfield  {journal} {\bibinfo  {journal} {Phys. Rev. E}\ }\textbf {\bibinfo
			{volume} {98}},\ \bibinfo {pages} {052124} (\bibinfo {year}
		{2018})}\BibitemShut {NoStop}%
	\bibitem [{\citenamefont {Kitaev}(2001{\natexlab{a}})}]{Kitaev_2001}%
	\BibitemOpen
	\bibfield  {author} {\bibinfo {author} {\bibfnamefont {A~.Y.}\ \bibnamefont
			{Kitaev}},\ }\bibfield  {title} {\enquote {\bibinfo {title} {Unpaired
				majorana fermions in quantum wires},}\ }\href {\doibase
		10.1070/1063-7869/44/10s/s29} {\bibfield  {journal} {\bibinfo  {journal}
			{Physics-Uspekhi}\ }\textbf {\bibinfo {volume} {44}},\ \bibinfo {pages}
		{131--136} (\bibinfo {year} {2001}{\natexlab{a}})}\BibitemShut {NoStop}%
	\bibitem [{\citenamefont {Kitaev}(2003)}]{KITAEV20032}%
	\BibitemOpen
	\bibfield  {author} {\bibinfo {author} {\bibfnamefont {A.~Y.}\ \bibnamefont
			{Kitaev}},\ }\bibfield  {title} {\enquote {\bibinfo {title} {Fault-tolerant
				quantum computation by anyons},}\ }\href {\doibase
		https://doi.org/10.1016/S0003-4916(02)00018-0} {\bibfield  {journal}
		{\bibinfo  {journal} {Ann. Phys.}\ }\textbf {\bibinfo {volume} {303}},\
		\bibinfo {pages} {2 -- 30} (\bibinfo {year} {2003})}\BibitemShut {NoStop}%
	\bibitem [{\citenamefont {Alicea}(2012)}]{Alicea_2012}%
	\BibitemOpen
	\bibfield  {author} {\bibinfo {author} {\bibfnamefont {J.}~\bibnamefont
			{Alicea}},\ }\bibfield  {title} {\enquote {\bibinfo {title} {New directions
				in the pursuit of majorana fermions in solid state systems},}\ }\href
	{\doibase 10.1088/0034-4885/75/7/076501} {\bibfield  {journal} {\bibinfo
			{journal} {Rep. Prog. Phys}\ }\textbf {\bibinfo {volume} {75}},\ \bibinfo
		{pages} {076501} (\bibinfo {year} {2012})}\BibitemShut {NoStop}%
	\bibitem [{\citenamefont {Molignini}\ \emph {et~al.}(2017)\citenamefont
		{Molignini}, \citenamefont {van Nieuwenburg},\ and\ \citenamefont
		{Chitra}}]{PhysRevB.96.125144}%
	\BibitemOpen
	\bibfield  {author} {\bibinfo {author} {\bibfnamefont {P.}~\bibnamefont
			{Molignini}}, \bibinfo {author} {\bibfnamefont {E.}~\bibnamefont {van
				Nieuwenburg}}, \ and\ \bibinfo {author} {\bibfnamefont {R.}~\bibnamefont
			{Chitra}},\ }\bibfield  {title} {\enquote {\bibinfo {title} {Sensing
				floquet-majorana fermions via heat transfer},}\ }\href {\doibase
		10.1103/PhysRevB.96.125144} {\bibfield  {journal} {\bibinfo  {journal} {Phys.
				Rev. B}\ }\textbf {\bibinfo {volume} {96}},\ \bibinfo {pages} {125144}
		(\bibinfo {year} {2017})}\BibitemShut {NoStop}%
	\bibitem [{\citenamefont {Elcock}\ and\ \citenamefont
		{Landsberg}(1957)}]{Elcock_1957}%
	\BibitemOpen
	\bibfield  {author} {\bibinfo {author} {\bibfnamefont {E~.W.}\ \bibnamefont
			{Elcock}}\ and\ \bibinfo {author} {\bibfnamefont {P.~T.}\ \bibnamefont
			{Landsberg}},\ }\bibfield  {title} {\enquote {\bibinfo {title} {Temperature
				dependent energy levels in statistical mechanics},}\ }\href {\doibase
		10.1088/0370-1301/70/2/301} {\bibfield  {journal} {\bibinfo  {journal} {Proc.
				Phys. Soc., B}\ }\textbf {\bibinfo {volume} {70}},\ \bibinfo {pages}
		{161--168} (\bibinfo {year} {1957})}\BibitemShut {NoStop}%
	\bibitem [{\citenamefont {Rushbrooke}(1940)}]{Rushbrooke}%
	\BibitemOpen
	\bibfield  {author} {\bibinfo {author} {\bibfnamefont {G.~S.}\ \bibnamefont
			{Rushbrooke}},\ }\bibfield  {title} {\enquote {\bibinfo {title} {On the
				statistical mechanics of assemblies whose energy-levels depend on the
				temperature},}\ }\href {\doibase 10.1039/TF9403601055} {\bibfield  {journal}
		{\bibinfo  {journal} {Trans. Faraday Soc.}\ }\textbf {\bibinfo {volume}
			{36}},\ \bibinfo {pages} {1055--1062} (\bibinfo {year} {1940})}\BibitemShut
	{NoStop}%
	\bibitem [{\citenamefont {de~Miguel}\ and\ \citenamefont
		{Rubi}(2016)}]{miguel1}%
	\BibitemOpen
	\bibfield  {author} {\bibinfo {author} {\bibfnamefont {R.}~\bibnamefont
			{de~Miguel}}\ and\ \bibinfo {author} {\bibfnamefont {J.~M.}\ \bibnamefont
			{Rubi}},\ }\bibfield  {title} {\enquote {\bibinfo {title} {Finite systems in
				a heat bath: Spectrum perturbations and thermodynamics},}\ }\href {\doibase
		10.1021/acs.jpcb.6b05591} {\bibfield  {journal} {\bibinfo  {journal} {J.
				Phys. Chem. B}\ }\textbf {\bibinfo {volume} {120}},\ \bibinfo {pages}
		{9180--9186} (\bibinfo {year} {2016})}\BibitemShut {NoStop}%
	\bibitem [{\citenamefont {de~Miguel}\ and\ \citenamefont
		{Rub{\'\i}}(2017)}]{miguel2}%
	\BibitemOpen
	\bibfield  {author} {\bibinfo {author} {\bibfnamefont {R.}~\bibnamefont
			{de~Miguel}}\ and\ \bibinfo {author} {\bibfnamefont {J.~M.}\ \bibnamefont
			{Rub{\'\i}}},\ }\bibfield  {title} {\enquote {\bibinfo {title}
			{Thermodynamics far from the thermodynamic limit},}\ }\href {\doibase
		10.1021/acs.jpcb.7b08621} {\bibfield  {journal} {\bibinfo  {journal} {J.
				Phys. Chem. B}\ }\textbf {\bibinfo {volume} {121}},\ \bibinfo {pages}
		{10429--10434} (\bibinfo {year} {2017})}\BibitemShut {NoStop}%
	\bibitem [{\citenamefont {Miguel}(2015)}]{miguel_temperature-dependent_2015}%
	\BibitemOpen
	\bibfield  {author} {\bibinfo {author} {\bibfnamefont {R.~de}\ \bibnamefont
			{Miguel}},\ }\bibfield  {title} {\enquote {\bibinfo {title}
			{Temperature-dependent energy levels and size-independent thermodynamics},}\
	}\href {\doibase 10.1039/C5CP02332G} {\bibfield  {journal} {\bibinfo
			{journal} {Phys. Chem. Chem. Phys.}\ }\textbf {\bibinfo {volume} {17}},\
		\bibinfo {pages} {15691--15693} (\bibinfo {year} {2015})}\BibitemShut
	{NoStop}%
	\bibitem [{\citenamefont {Yamano}(2016)}]{yamano2}%
	\BibitemOpen
	\bibfield  {author} {\bibinfo {author} {\bibfnamefont {T.}~\bibnamefont
			{Yamano}},\ }\bibfield  {title} {\enquote {\bibinfo {title} {Efficiencies of
				thermodynamics when temperature-dependent energy levels exist},}\ }\href
	{\doibase 10.1039/C5CP07572F} {\bibfield  {journal} {\bibinfo  {journal}
			{Phys. Chem. Chem. Phys.}\ }\textbf {\bibinfo {volume} {18}},\ \bibinfo
		{pages} {7011--7014} (\bibinfo {year} {2016})}\BibitemShut {NoStop}%
	\bibitem [{\citenamefont {Yamano}(2017)}]{yamano_effect_2017}%
	\BibitemOpen
	\bibfield  {author} {\bibinfo {author} {\bibfnamefont {T.}~\bibnamefont
			{Yamano}},\ }\bibfield  {title} {\enquote {\bibinfo {title} {Effect of
				temperature-dependent energy levels on exergy},}\ }\href {\doibase
		10.1088/2399-6528/aa95e4} {\bibfield  {journal} {\bibinfo  {journal} {J.
				Phys. Commun.}\ }\textbf {\bibinfo {volume} {1}},\ \bibinfo {pages} {055007}
		(\bibinfo {year} {2017})}\BibitemShut {NoStop}%
	\bibitem [{\citenamefont {Shental}\ and\ \citenamefont
		{Kanter}(2009)}]{Shental_2009}%
	\BibitemOpen
	\bibfield  {author} {\bibinfo {author} {\bibfnamefont {O.}~\bibnamefont
			{Shental}}\ and\ \bibinfo {author} {\bibfnamefont {I.}~\bibnamefont
			{Kanter}},\ }\bibfield  {title} {\enquote {\bibinfo {title} {Shannon meets
				carnot: Generalized second thermodynamic law},}\ }\href {\doibase
		10.1209/0295-5075/85/10006} {\bibfield  {journal} {\bibinfo  {journal} {{EPL}
				(Europhysics Letters)}\ }\textbf {\bibinfo {volume} {85}},\ \bibinfo {pages}
		{10006} (\bibinfo {year} {2009})}\BibitemShut {NoStop}%
	\bibitem [{\citenamefont {Lebowitz}\ and\ \citenamefont
		{Percus}(1961)}]{lebowitz_thermodynamic_1961}%
	\BibitemOpen
	\bibfield  {author} {\bibinfo {author} {\bibfnamefont {J.~L.}\ \bibnamefont
			{Lebowitz}}\ and\ \bibinfo {author} {\bibfnamefont {J.~K.}\ \bibnamefont
			{Percus}},\ }\bibfield  {title} {\enquote {\bibinfo {title} {Thermodynamic
				{Properties} of {Small} {Systems}},}\ }\href {\doibase
		10.1103/PhysRev.124.1673} {\bibfield  {journal} {\bibinfo  {journal} {Phys.
				Rev.}\ }\textbf {\bibinfo {volume} {124}},\ \bibinfo {pages} {1673--1681}
		(\bibinfo {year} {1961})}\BibitemShut {NoStop}%
	\bibitem [{\citenamefont {Aydin}\ and\ \citenamefont {Sisman}(2019)}]{sisman}%
	\BibitemOpen
	\bibfield  {author} {\bibinfo {author} {\bibfnamefont {A.}~\bibnamefont
			{Aydin}}\ and\ \bibinfo {author} {\bibfnamefont {A.}~\bibnamefont {Sisman}},\
	}\bibfield  {title} {\enquote {\bibinfo {title} {Quantum shape effects and
				novel thermodynamic behaviors at nanoscale},}\ }\href {\doibase
		https://doi.org/10.1016/j.physleta.2019.01.009} {\bibfield  {journal}
		{\bibinfo  {journal} {Phys. Lett. A}\ }\textbf {\bibinfo {volume} {383}},\
		\bibinfo {pages} {655 -- 665} (\bibinfo {year} {2019})}\BibitemShut {NoStop}%
	\bibitem [{\citenamefont {Bedeaux}\ and\ \citenamefont
		{Kjelstrup}(2018)}]{bedeaux_hills_2018}%
	\BibitemOpen
	\bibfield  {author} {\bibinfo {author} {\bibfnamefont {D.}~\bibnamefont
			{Bedeaux}}\ and\ \bibinfo {author} {\bibfnamefont {S.}~\bibnamefont
			{Kjelstrup}},\ }\bibfield  {title} {\enquote {\bibinfo {title} {Hill's
				nano-thermodynamics is equivalent with {Gibbs}' thermodynamics for surfaces
				of constant curvatures},}\ }\href {\doibase 10.1016/j.cplett.2018.07.031}
	{\bibfield  {journal} {\bibinfo  {journal} {Chem. Phys. Lett.}\ }\textbf
		{\bibinfo {volume} {707}},\ \bibinfo {pages} {40--43} (\bibinfo {year}
		{2018})}\BibitemShut {NoStop}%
	\bibitem [{\citenamefont {Varshni}(1967)}]{VARSHNI1967}%
	\BibitemOpen
	\bibfield  {author} {\bibinfo {author} {\bibfnamefont {Y.P.}\ \bibnamefont
			{Varshni}},\ }\bibfield  {title} {\enquote {\bibinfo {title} {Temperature
				dependence of the energy gap in semiconductors},}\ }\href {\doibase
		https://doi.org/10.1016/0031-8914(67)90062-6} {\bibfield  {journal} {\bibinfo
			{journal} {Physica}\ }\textbf {\bibinfo {volume} {34}},\ \bibinfo {pages}
		{149 -- 154} (\bibinfo {year} {1967})}\BibitemShut {NoStop}%
	\bibitem [{\citenamefont {Emin}(1984)}]{PhysRevB.30.5766}%
	\BibitemOpen
	\bibfield  {author} {\bibinfo {author} {\bibfnamefont {D.}~\bibnamefont
			{Emin}},\ }\bibfield  {title} {\enquote {\bibinfo {title} {Effect of
				temperature-dependent energy-level shifts on a semiconductor's peltier
				heat},}\ }\href {\doibase 10.1103/PhysRevB.30.5766} {\bibfield  {journal}
		{\bibinfo  {journal} {Phys. Rev. B}\ }\textbf {\bibinfo {volume} {30}},\
		\bibinfo {pages} {5766--5770} (\bibinfo {year} {1984})}\BibitemShut {NoStop}%
	\bibitem [{\citenamefont {Allen}\ and\ \citenamefont
		{Heine}(1976)}]{Allen_1976}%
	\BibitemOpen
	\bibfield  {author} {\bibinfo {author} {\bibfnamefont {P.~B.}\ \bibnamefont
			{Allen}}\ and\ \bibinfo {author} {\bibfnamefont {V.}~\bibnamefont {Heine}},\
	}\bibfield  {title} {\enquote {\bibinfo {title} {Theory of the temperature
				dependence of electronic band structures},}\ }\href {\doibase
		10.1088/0022-3719/9/12/013} {\bibfield  {journal} {\bibinfo  {journal} {J.
				Phys. C: Solid State Phys.}\ }\textbf {\bibinfo {volume} {9}},\ \bibinfo
		{pages} {2305--2312} (\bibinfo {year} {1976})}\BibitemShut {NoStop}%
	\bibitem [{\citenamefont {Patrick}\ and\ \citenamefont
		{Giustino}(2014)}]{Patrick_2014}%
	\BibitemOpen
	\bibfield  {author} {\bibinfo {author} {\bibfnamefont {C.~E.}\ \bibnamefont
			{Patrick}}\ and\ \bibinfo {author} {\bibfnamefont {F.}~\bibnamefont
			{Giustino}},\ }\bibfield  {title} {\enquote {\bibinfo {title} {Unified theory
				of electron{\textendash}phonon renormalization and phonon-assisted optical
				absorption},}\ }\href {\doibase 10.1088/0953-8984/26/36/365503} {\bibfield
		{journal} {\bibinfo  {journal} {J. Phys.: Condens. Matter}\ }\textbf
		{\bibinfo {volume} {26}},\ \bibinfo {pages} {365503} (\bibinfo {year}
		{2014})}\BibitemShut {NoStop}%
	\bibitem [{\citenamefont {Dykman}\ \emph {et~al.}(2017)\citenamefont {Dykman},
		\citenamefont {Kono}, \citenamefont {Konstantinov},\ and\ \citenamefont
		{Lea}}]{PhysRevLett.119.256802}%
	\BibitemOpen
	\bibfield  {author} {\bibinfo {author} {\bibfnamefont {M.~I.}\ \bibnamefont
			{Dykman}}, \bibinfo {author} {\bibfnamefont {K.}~\bibnamefont {Kono}},
		\bibinfo {author} {\bibfnamefont {D.}~\bibnamefont {Konstantinov}}, \ and\
		\bibinfo {author} {\bibfnamefont {M.~J.}\ \bibnamefont {Lea}},\ }\bibfield
	{title} {\enquote {\bibinfo {title} {Ripplonic lamb shift for electrons on
				liquid helium},}\ }\href {\doibase 10.1103/PhysRevLett.119.256802} {\bibfield
		{journal} {\bibinfo  {journal} {Phys. Rev. Lett.}\ }\textbf {\bibinfo
			{volume} {119}},\ \bibinfo {pages} {256802} (\bibinfo {year}
		{2017})}\BibitemShut {NoStop}%
	\bibitem [{\citenamefont {Kol{\'a}{\v r}}\ \emph {et~al.}(2017)\citenamefont
		{Kol{\'a}{\v r}}, \citenamefont {Ryabov},\ and\ \citenamefont
		{Filip}}]{kolar_optomechanical_2017}%
	\BibitemOpen
	\bibfield  {author} {\bibinfo {author} {\bibfnamefont {M.}~\bibnamefont
			{Kol{\'a}{\v r}}}, \bibinfo {author} {\bibfnamefont {A.}~\bibnamefont
			{Ryabov}}, \ and\ \bibinfo {author} {\bibfnamefont {R.}~\bibnamefont
			{Filip}},\ }\bibfield  {title} {\enquote {\bibinfo {title} {Optomechanical
				oscillator controlled by variation in its heat bath temperature},}\ }\href
	{\doibase 10.1103/PhysRevA.95.042105} {\bibfield  {journal} {\bibinfo
			{journal} {Phys. Rev. A}\ }\textbf {\bibinfo {volume} {95}},\ \bibinfo
		{pages} {042105} (\bibinfo {year} {2017})}\BibitemShut {NoStop}%
	\bibitem [{\citenamefont {Seifert}(2016)}]{seifert_first_2016}%
	\BibitemOpen
	\bibfield  {author} {\bibinfo {author} {\bibfnamefont {U.}~\bibnamefont
			{Seifert}},\ }\bibfield  {title} {\enquote {\bibinfo {title} {First and
				{Second} {Law} of {Thermodynamics} at {Strong} {Coupling}},}\ }\href
	{\doibase 10.1103/PhysRevLett.116.020601} {\bibfield  {journal} {\bibinfo
			{journal} {Phys. Rev. Lett.}\ }\textbf {\bibinfo {volume} {116}},\ \bibinfo
		{pages} {020601} (\bibinfo {year} {2016})}\BibitemShut {NoStop}%
	\bibitem [{\citenamefont {Talkner}\ and\ \citenamefont
		{H{\"a}nggi}(2016)}]{talkner_open_2016}%
	\BibitemOpen
	\bibfield  {author} {\bibinfo {author} {\bibfnamefont {P.}~\bibnamefont
			{Talkner}}\ and\ \bibinfo {author} {\bibfnamefont {P.}~\bibnamefont
			{H{\"a}nggi}},\ }\bibfield  {title} {\enquote {\bibinfo {title} {Open system
				trajectories specify fluctuating work but not heat},}\ }\href {\doibase
		10.1103/PhysRevE.94.022143} {\bibfield  {journal} {\bibinfo  {journal} {Phys.
				Rev. E}\ }\textbf {\bibinfo {volume} {94}},\ \bibinfo {pages} {022143}
		(\bibinfo {year} {2016})}\BibitemShut {NoStop}%
	\bibitem [{\citenamefont {Kitaev}(2001{\natexlab{b}})}]{kitaev}%
	\BibitemOpen
	\bibfield  {author} {\bibinfo {author} {\bibfnamefont {A.~Y.}\ \bibnamefont
			{Kitaev}},\ }\bibfield  {title} {\enquote {\bibinfo {title} {Unpaired
				{{Majorana}} fermions in quantum wires},}\ }\href {\doibase
		10.1070/1063-7869/44/10S/S29} {\bibfield  {journal} {\bibinfo  {journal}
			{Physics-Uspekhi}\ }\textbf {\bibinfo {volume} {44}},\ \bibinfo {pages}
		{131--136} (\bibinfo {year} {2001}{\natexlab{b}})}\BibitemShut {NoStop}%
\end{thebibliography}

%

\end{document}